\documentclass[]{aastex}

\usepackage{emulateapj5}
\usepackage{onecolfloat}
\usepackage{graphicx}
\usepackage{fancyheadings}
\usepackage{ulem}
\usepackage{rotating}
\usepackage{lscape}

\def\DLAs{damped {\lya} systems}
\def\DLA{damped {\lya} system}

\def\etal{et al.}

\def\lya{Ly$\alpha$ }
\def\smpy{M$_{\odot}{\rm \ yr^{-1} }$}
\def\smpykpc{M$_{\odot}{\rm \ yr^{-1} \ kpc^{-2}}$}

\def\lamcii{${\Lambda}_{\rm CII}$}

\def\kms{km~s$^{-1}$ }

\def\micron{$\mu$m}

\def\ncnm{$n_{\rm CNM}$}

\def\cm2{\, \rm cm^{-2}}
\def\cm3{cm$^{-3}$}

\def\cmma{\;\;\; ,}

\def\ps{$\dot{\psi_{*}}$}

\def\ciis{C II$^{*}$}
\def\nh{$N$(H I)}
\def\bc{$b_{\rm C}$}
\def\nha{$N^{a}$(H I)}
\def\lclos{$l_{c}$}
\def\lcobs{$l_{c}^{obs}$}
\def\lcpeak{$l_{c}^{peak}$}

\def\lcn{$l_{c}$($n$)}

\def\jnu{$J_{\nu}$}
\def\jnuphot{$J_{\nu}^{phot}$}
\def\jnustar{$J_{\nu}^{stars}$}
\def\jnubkd{$J_{\nu}^{bkd}$}
\def\jnucstar{$J_{\nu}^{{\rm C II^{*}}}$}
\def\hnu{$h{\nu}$}
\def\hnuH{$h{\nu}_{H}$}
\def\jnuH{$J_{\nu_{H}}$}

\def\junit{ergs cm$^{-2}$ s$^{-1}$ Hz$^{-1}$ sr$^{-1}$}
\def\nh{$N$(H I)}
\def\teno{2${\times}$10$^{20}$ cm$^{-2}$}

\begin{document}

\twocolumn[%
\submitted{Accepted for Publication by the  Astrophysical Journal July 14, 2004}

\title{ON THE NATURE OF THE HEAT SOURCE FOR DAMPED \lya\ SYSTEMS}

\author{ ARTHUR M. WOLFE \altaffilmark{1} \& J. CHRISTOPHER HOWK \altaffilmark{1}\\
Department of Physics and Center for Astrophysics and Space Sciences; \\
University of California, San
Diego; \\
C--0424; La Jolla, CA 92093\\
{\bf awolfe@ucsd.edu, howk@ucsd.edu}}

\author{}

\author{ ERIC GAWISER\\
Department of Astronomy; \\
Yale University \\
P.O. Box 208101 \\
New Haven, CT 06520-8101\\
{\bf egawiser@astro.yale.edu}}

\author{}

\author{JASON X. PROCHASKA\altaffilmark{1}\\
UCO-Lick Observatory; \\
University of California, Santa Cruz\\
Santa Cruz, CA; 95464\\
{\bf xavier@ucolick.org}}

\author{and}

\author{SEBASTIAN LOPEZ \\
Universidad de Chile; \\
Departamento de Astronomia\\
Casilla 36-D, Las Condes,\\
Santiago, Chile\\
{\bf slopez@das.uchile.cl}}

\begin{abstract}

We investigate the heat source of the neutral gas comprising
{\DLAs}.
Unlike the {\lya} forest, where the extragalactic background
radiation field ionizes and heats the gas,
we find that grain photoelectric heating by the FUV
background is not sufficient to balance the 
C II 158 {$\mu$m} cooling rate inferred from {\DLAs}
%where {\ciis} absorption is detected.
In these systems, a local
energy source is required.
We show that in the case of
the $z$=1.919 {\DLA} toward Q2206$-$19, the local source is
FUV emission from the associated galaxy found by M$\o$ller
{\etal} (2002):
the mean intensity
inferred from photometry
is in good agreement with the intensity {\jnustar} required to
explain the cooling rate.
The FUV mean intensity predicted for a cold neutral medium (CNM) model,
{\jnustar}=(1.7$^{+2.7}_{-1.0}$){$\times$}10$^{-18}${\junit} (95$\%$ c.l.),
is the largest expected from our {\ciis}
study of 45 {\DLAs}.This may explain why this is the
only confirmed {\DLA} yet detected in emission at $z$ $>$ 1.9.
We argue that in most {\DLAs} with detected  {\ciis} absorption,
{\jnustar} is between 10$^{-19}$ and  10$^{-18}$ {\junit} and heats
the gas which is a CNM.  By contrast, in most {\DLAs} with
upper limits on {\ciis} absorption the gas is a warm neutral
medium (WNM).
Surprisingly, the upper limits are compatible with
the same range of {\jnustar} values
suggesting the majority of {\DLAs} are heated by radiation fields
generated by a limited range
of star formation rates per unit H I area, between 10$^{-3}$ and 10$^{-2}$.
{\smpykpc}.
We also show that {\ciis} absorption is unlikely to arise in
gas that is ionized.

 \end{abstract}


]
\altaffiltext{1}{Visiting Astronomer, W.M. Keck Telescope.
The Keck Observatory is a joint facility of the University
of California and the California Institute of Technology.}

\pagestyle{fancyplain}
\lhead[\fancyplain{}{\thepage}]{\fancyplain{}{Wolfe, Prochaska, \& Gawiser}}
\rhead[\fancyplain{}{STAR FORMATION IN DAMPED {\lya} SYSTEMS}]{\fancyplain{}{\thepage}}
\setlength{\headrulewidth=0pt}
\cfoot{}

\section{INTRODUCTION}

External background radiation is the widely accepted 
heat source for the {\lya}
forest.
Defined to have
H I
column densities
{\nh} $<$ 10$^{17}$ cm$^{-2}$ 
the 
{\lya} 
forest clouds
are optically thin at the Lyman limit. As a result,
ionizing background radiation, i.e., radiation with
photon energies $h{\nu} > h{\nu}_{H}$ = 13.6 eV, 
penetrates the interior of the clouds,
ionizing them such that
the neutral gas fraction $n({\rm H^{0}})/n$$<$10$^{-5}$, 
and heats them to  
temperatures in excess of 10000 K 
(Rauch 1998).
Indeed,
several authors have constructed scenarios within
the context of CDM cosmogonies in which  most of the baryons
in the universe at $z$$\sim$3 reside in the {\lya} forest and are 
photoionized by background radiation (e.g., Miralda-Escud$\rm \acute e$
{\etal} 1996; Machacek {\etal} 2000).

Calculations by Haardt \& Madau (1996; 2003)
show that radiation emitted from the integrated
population of QSOs and galaxies results in a mean 
background intensity $J_{\nu_H}$
$\sim$ 10$^{-21}$ {\junit} at $z$ $\sim$ 3.
Independent tests,
such as the decrease with redshift of the number of absorbers 
per unit redshift in the
vicinity of the QSO, i.e.,  
``the proximity effect'', indicate similar
values for {\jnuH} (e.g., Scott {\etal} 2002). 
Ionizing background radiation has also been invoked to
explain the ionization structure of metals in
the {\lya} forest, and in higher column-density
QSO absorption systems 
selected for C IV ${\lambda}{\lambda}$ 1548, 1551 absorption (e.g. Mo
\& Miralda-Escud$\rm \acute e$ 1995),
Lyman limit absorption (Prochaska 1999),
and Mg II ${\lambda}{\lambda}$ 2796, 2803
absorption (e.g. Churchill, Vogt, \& Charlton  2003). 
The predicted temperatures in these absorbers are
also about 10000 K and the neutral gas fraction, $n({\rm H^{0}})/n$ $<<$ 1.

In this paper  we consider the nature of
the heat source for
another
class of QSO absorption systems, the {\DLAs} (see
Storrie-Lombardi \& Wolfe 2000).
While the {\DLAs}
are subject to the same external radiation fields as the other
absorbers, they differ in one important respect: the gas is
neutral (e.g. Vladilo {\etal} 2001; Prochaska {\etal} 2002). 
Defined to have H I column densities {\nh}{$\ge$}{\teno},
{\DLAs}  have
Lyman-limit optical depths,  $\tau_{{\nu_{H}}}$
$\ge$1000. As a result,
%ionizing photons with
%energies $h{\nu_{H}}$$\le$$h{\nu}$$\le$0.4 keV, i.e.,
the same photons that ionize and heat the {\lya} forest 
do not
penetrate the interior of the {\DLAs}.
Rather, these gas layers are penetrated only 
by background radiation with photon energies {\hnu} $\gtrsim$ 400 eV 
and $h{\nu}$ $<$ $h{\nu_{H}}$. 
While the low-energy FUV photons do not ionize H, they do ionize
neutral species of atoms with ionization potentials
less than {\hnuH} such as C, Fe, and Zn, and more
importantly, could contribute to the heat input
of {\DLAs} through grain photoelectric heating
(Bakes \& Tielens 1994). Similarly,  
X-ray photons can also heat
the gas though photoionization.

Does such background radiation dominate the heat input to 
the {\DLAs}?
We shall address this question
empirically
since
cooling rates have recently been measured for a representative
sample of $\sim$ 50 {\DLAs} (see Wolfe, Prochaska, \& Gawiser 2003,
hereafter referred to as  WPG, for a discussion of the
original sample of 33 {\DLAs}). 
The answer has important
implications. 
If heating by external sources does not match
the measured cooling rates,
then internal sources of heating must be found. This 
would distinguish {\DLAs} from all other classes
of QSO absorption systems, which are plausibly heated
by background radiation. 

This paper is organized as follows.
In $\S$ 2 we describe our technique for 
obtaining the thermal equilibria of 
{\DLAs} heated by external background
radiation. In particular we   compute the
thermal equilibria of neutral gas layers exposed to
background radiation fields recently computed by
Haardt \& Madau (2003). These authors computed
backgrounds
originating from the integrated population of
(1) QSOs alone, 
and (2) QSOs 
and galaxies. To maximize the heating rates 
we consider those backgrounds that include 
galaxies because
the galaxy contribution dominates at 
$h{\nu}$$\le$$h{\nu_{H}}$, where
FUV radiation (6 $<$ $h{\nu}$ $<$ 13.6 eV) heats the gas by the
grain photoelectric mechanism (WPG). This is the same
process by which the neutral ISM of the Galaxy is 
heated (e.g. Bakes \& Tielens 1994), and it
is relevant here given the evidence 
for dust in {\DLAs} (Pettini {\etal} 1994;
Pei \& Fall 1995; WPG). At $h{\nu}$ $\gtrsim$ 400 eV,
soft X-rays heat the gas by
photoionizing H, He,
and abundant elements.
X-ray heating occurs primarily through  the photoionization of H and He
because of the low metallicities of {\DLAs}.
In $\S$3 we apply these results
to
test
the background heating hypothesis. We
compare
the predicted
heating rates to the 
158 {\micron} cooling
rates deduced from the strength of {\ciis} $\lambda$ 1335.7
and damped {\lya}  absorption
(see WPG). We study the multi-phase structure
of the $z$ = 1.919 {\DLA} toward Q2206$-$19,
and then discuss results for the full sample
of 45 {\DLAs}. 
%the predicted
%heating rates are too low in those
%{\DLAs} with detected {\ciis}  absorption. 
%On the other hand, we cannot rule out
%background heating as the dominant heat input for 
%{\DLAs} in which upper limits on
%{\ciis} absorption have been set. 
In $\S$ 4
we examine the nature of possible internal heat
sources by focusing on the $z$=1.919 {\DLA}
toward Q2206$-$19. This {\DLA} is ideal  for such a 
study because the presence of an
associated galaxy at the same redshift, which is detected
via its rest-frame FUV emission
(M{$\o$}ller {\etal} 2002), allows one to make an independent
estimate of the FUV radiation field incident on the
absorbing gas. Since Fe is believed to be depleted
in this DLA, the radiation is incident
on dust, and the result is grain photoelectric
heating. WPG show how to deduce 
the local FUV radiation
field from the strength of {\ciis} absorption.
In $\S$ 5 we investigate the energy input for all the
{\DLAs} in our sample, i.e., those with positive detections
and those with upper
limits on {\ciis} absorption, to determine whether
they are heated by background
radiation alone or require internal heat sources.
In $\S$ 6 we discuss the possibility that {\ciis} absorption
arises in ionized gas.
Finally, concluding remarks are given
in $\S$ 7. 

%absorption and the Fe to Si ratio. 
%We show the FUV mean intensity
%deduced by this technique to be consistent
%with the FUV radiation field emitted by
%the associated galaxy.
%Therefore, the gas in this {\DLA} is likely heated
%by radiative feedback rather than mechanical or some
%other type of radiationless process.
%When 
%combined with other correlations discussed by
%Wolfe, Gawiser, \& Prochaska (2003; hereafter referred to as WGP),
%%we find
%this evidence supports the hypothesis
%that {\DLAs} are heated by the grain photoelectric
%mechanism, and that the heat source is internally generated starlight 
%(see WPG). Since FUV radiation is emitted primarily
%%by short-lived stars on the upper main sequence,
%we conclude that star formation
%is an ongoing process in {\DLAs}.

Throughout this paper we adopt a cosmology
consistent with the WMAP (Bennett {\etal} 2003) results,
($\Omega_{m}$, $\Omega_{\Lambda}$, $h$) = (0.3, 0.7, 0.7).

%This paper is organized as follows: In $\S$ 2 we 
%compute heating rates due to background
%X-ray and FUV radiation. In $\S$ 3 
%and photoionization by soft X-rays for a representative sample
%of {\DLAs}. We show that neither process is adequate for balancing
%the observed [C II] 158 {\micron} cooling rates. In $\S$ 3
%heating by FUV radiation emitted by massive stars. We then
%test this hypothesis by evaluating the cooling rate for a {\DLA}
%observed to be associated with a galaxy emitting FUV radiation.
%We show the star formation rate implied by
%the heating rate is consistent with
%the star formation rate indicated by the emitted FUV radiation
%field. $\S$ 4 describes some
%of the implications of this conclusion regarding the two-phase
%structure of the absorbing gas and the connection between
%metal production and star formation 
%in {\DLAs}.

\section{THERMAL EQUILIBRIA OF BACKGROUND HEATED DAMPED {\lya} SYSTEMS}

Consider a slab of neutral gas exposed to an isotropic background radiation
field with mean intensity $J_{\nu}$. We first compute the heating
rates due to soft X-rays 
and FUV radiation. We then calculate the thermal equilibria resulting
from a balance between heating and cooling. 
%The resulting 
%[C II] 158 {\micron} emission rates per H atom
%are compared with observations.

\subsection{Heating}

Wolfire {\etal} (1995; hereafter W95) find the primary ionization rate 
of atomic species $i$ due to soft X-rays to be

\begin{equation}
 {\zeta^{i}_{XR}} = 4{\pi}{\int {J_{\nu} \over h{\nu}}{\rm exp}[-{\sigma_{\nu}}N^{a}({\rm H I})]{\sigma_{\nu}}^{i}d{\nu}} \ {\rm \ s^{-1}}
\cmma
\label{eq:zetXRp}
\end{equation}

\noindent where {$\sigma_{\nu}^{i}$} is the photoionization cross-section
per H atom of species $i$ and the factor  {$\sigma_{\nu}${\nha}} is the optical
depth due to H, He, and abundant metals such as
C, O, Fe, etc. in an attenuating
column density {\nha}.
To simulate conditions in the midplane of
the gas layer
we let {\nha} =
0.5{\nh},
where {\nh} is the H I column density detected along the
line of sight.
We compute the total ionization rate of H and He,
including secondary ionizations, using the prescription of W95
including their modifications of the results
of Shull \& van-Steenberg (1985). The resulting heating rate is given by 

\begin{equation}
 {\Gamma_{XR}} = 4{\pi}{\sum\limits_{i}}{\int {J_{\nu} \over h{\nu}}{\rm exp}[-{\sigma_{\nu}}N^{a}({\rm H I})]{\sigma_{\nu}}^{i}E_{h}(E^{i},x)d{\nu}} 
\cmma
\label{eq:GammXR}
\end{equation}

\noindent where the summation is over species suffering primary ionization,
and $E_{h}(E^{i},x)$ is the energy deposited as heat by a primary electron
originating from atomic species $i$ with energy $E^{i}$ in gas with electron
fraction $x$. We used the results of Shull \& van-Steenberg (1985)
to  compute $E_{h}$, and an updated version
of the Morrison \& McCammon (1983) results to compute
$\sigma_{\nu}^{i}$ and $\sigma_{\nu}$ (McCammon 2003).

\begin{table} \small
\begin{center}
\caption{Parameters for Grain Heating Models} \label{data}
\begin{tabular}{lccccc}
Name &[C/H]$^{a}$&$\kappa \ ^{b}$ &{\bc} \ $^{c}$& $R_{V} \ ^{d}$&Reference$^{e}$ \\
\tableline
WD-low&[Si/H]$-$0.2 & $^{f}$   & 0& 3.1&(1)   \\
WD-high&[Si/H] & $^{g}$&6{$\times$}10$^{-5}$& 3.1&(1)   \\
BT &[Si/H]&$^{g}$ &1.26{$\times$}10$^{-6}$ &...& (2)  \\

\end{tabular}
\end{center}
\tablenotetext{a}{Carbon abundance relative to solar.}
\tablenotetext{b}{dust-to-gas ratio relative to the Milky Way ISM}
\tablenotetext{c}{Abundance of C atoms per interstellar H nucleus in PAHs}
\tablenotetext{d}{$R_{V}$$\equiv$ $A_{V}$/$E_{B-V}$, where $A_{V}$ is visual extinction}
\tablenotetext{}{ and $E_{B-V}$ is the
color excess.}
\tablenotetext{e}{(1) Weingartner \& Draine 2001; (2) Bakes \& Tielens
1994}
\tablenotetext{f}{10$^{[{\rm Si/H}]}$(10$^{-0.2}-10^{[{\rm Fe/Si}]}$)}
\tablenotetext{g}{10$^{[{\rm Si/H}]}$(1$-$10$^{[{\rm Fe/Si}]}$)}
\end{table}

To compute the grain photoelectric heating rate due to FUV
radiation we used the Weingartner
\& Draine (2001)
expression

\begin{equation}
{\Gamma_{d}} = 10^{-26}{\kappa}G_{0}f(G_{0}{\sqrt T}/n_{e}, c_{i}){\rm exp}(-{\tau^{d}_{\nu}})
\cmma
\label{eq:Gammpe}
\end{equation}

%\begin{equation}
%{\Gamma_{pe}} = 10^{-26}{\kappa}(G{\rm exp}[-{\tau^{d}_{\nu}}]){{C_{0}+C_{1}T^{C_{4}}} \over {1+C_{2}(G{\sqrt T}/n_{e})^{C_{5}}{\Biggl [}1+C_{3}(G{\sqrt T}/n_{e})^{C_{6}}{\Biggr ]}}} {\rm \ ergs \ s^{-1} \ H^{-1}}
%\cmma
%\label{eq:Gammpe}
%\end{equation}

\noindent where {$\Gamma_{d}$} is in
units of ergs s$^{-1}$ H$^{-1}$, $\kappa$ is the dust-to-gas ratio of the {\DLA} relative
to that of the ISM (ISM refers to
the ISM of the Milky Way Galaxy) , $G_{0}$ is 4{$\pi$}$\int J_{\nu}d{\nu}$ 
integrated between 6 and 13.6 eV and is in
units of 1.6{$\times$}10$^{-3}$ ergs s$^{-1}$ cm$^{-2}$, 
$f$, which is proportional to the heating efficiency, is a function of
{$G_{0}{\sqrt T}/n_{e}$} and the fitting constants, $c_{i}$ (where $i$=0$\rightarrow$6),
and $\tau^{d}_{\nu}$  
is the dust optical depth corresponding
to $N^{a}$(H I). 

Weingartner \& Draine (2001) show that
$\Gamma_{d}$ is especially sensitive to two quantities:
grain composition, i.e. the fraction of dust in silicate
and carbonaceous grains, and the fraction of dust in
small grains (radius $<$ 15 {\AA}).
Both are crucial because 
computations carried out so far show that small carbonaceous grains
are the most efficient sources of photoelectric heating.
The detection of the 2175 {\AA} absorption feature in two
{\DLAs} provide evidence for small carbonaceous grains
at $z$ $<$ 1 (Junkkarinen  {\etal} 2004;  Motta {\etal} 2002),
but the evidence is less clear for the redshift range of our
sample. 
Pei, Fall, \& Bechtold (1991) found no evidence for the 2175 {\AA}
feature, yet detection was expected based on their reddening model. 
Thus {\DLAs} with $z$ $>$ 2 could be comprised of silicate
dust alone. However, all the published models for grain
photoelectric heating include a population of small
carbonaceous grains.

To address this problem we consider models with
a range of heating efficiencies that should include
that of silicate dust. We adopt a 
model in which 
the fraction of carbon locked in PAHs 
per H nucleus, {\bc}=0, and 
extinction ratio $R_{V}$=3.1, 
which is valid for the
type of diffuse gas expected in {\DLAs} (see Table 2, row
19 in Weingartner \& Draine 2001);
where $R_{V}$$\equiv$$A_{V}$/$E_{B-V}$,  $A_{V}$ is the
visual extinction, and $E_{B-V}$ is the color excess.
We also assume
a radiation
spectrum approximating Draine's (1978) fit
to the interstellar radiation field in which
case $G_{0}$$\approx$$J_{\nu}$/(10$^{-19}$ {\junit}) 
at $\lambda$ = 1500 {\AA} (see WPG). Application
of this approximation to background radiation 
involves no loss in generality
since the background and interstellar radiation fields at
{\hnu} $<$ {\hnuH}
arise from similar stellar populations (e.g. Bruzual \& Charlot 2003).
The resulting
heating efficiency is the lowest  predicted for the
Weingartner \& Draine (2001) models.
The highest heating efficiency is predicted 
for their
model with $R_{V}$=3.1, {\bc}=6{$\times$}10$^{-5}$, and an
interstellar radiation spectrum (see their Table 2,
last row). We shall examine the following three dust
compositions: (1) a mix of silicate and carbonaceous
grains in which {\bc}=0 and
$R_{V}$=3.1,
(2) a mix of silicate and carbonaceous grains in which 
{\bc}=6{$\times$}10$^{-5}$
and $R_{V}$=3.1,
and (3) 
carbonaceous grains alone within the
heating model of Bakes \& Tielens (1994). In this way we
shall  test the
sensitivity of our results to changes in grain properties.
We show below that {\em our results concerning the plausibility
of background heating are independent of the properties
of the grains, and depend only on the carbon abundance
and assumptions about thermal equilibrium.}

Next we use the WPG method for computing $\kappa$. Our prescription
computes the fraction of Fe in grains and then assumes the number of depleted
C or Si atoms per depleted Fe atom to be the same in {\DLAs}
as in the Galaxy ISM. The depletion of Fe is calculated from 
the ratio of the 
observed amount of Fe to an undepleted element: Zn or S would
be ideal for this purpose.  WPG use Si because (i) Si
is measured over a wider range of redshifts than Zn,
(ii) S is difficult to measure since the observed S II transitions
occur frequently in the {\lya} forest, and (iii) though Si is undoubtedly
depleted, the observed depletion level is very
low; i.e., [Si/S]$\approx$0. 
\footnote {The abundance ratio of elements X and Y is
given by
[X/Y]=log$_{10}$(X/Y)$-$log$_{10}$(X/Y)$_{\odot}$.}
This does not lead to inconsistencies
in the case of silicate dust because the computed values of
{$\kappa$} are consistent with
Si being depleted typically by less than 20 $\%$.
WPG find
${\kappa} = 10^{[{\rm Si/H}]_{int}}{\Bigl (}10^{{\rm [Fe/Si]}_{int}}-10^{[\rm Fe/Si]_{gas}}{\Bigr )}$
where [Si/H]$_{int}$ is the intrinsic abundance of Si relative
to H, [Fe/Si]$_{int}$
is the intrinsic abundance of Fe relative to Si, and [Fe/Si]$_{gas}$
is the measured abundance of Fe relative to Si. 
Following  WPG we assume minimal depletion occurs when
[Fe/Si]$_{int}$=$-$0.2, i.e., an intrinsic enhancement of
$\alpha$ elements such as Si,
and maximum depletion  occurs
when 
[Fe/Si]$_{int}$=0. Therefore, we adopt two Weingartner
\& Draine (2001) models, which should
bracket a plausible range in heating rates. The ``WD-low''
model combines dust composition (1) with  minimal depletion: this
model is similar to one of the ``SMC'' models
considered by WPG and results in the lowest heating rates. 
\footnote{Minimal depletion does not
apply to {\DLAs}  in which [Fe/Si]$_{gas}$ $>$ $-$0.2.
%minimal depletion does not apply 
%the {\DLAs} toward Q0201$+$11,
%Q1021$+$30, Q1506$+$52, Q2241$+$13, and Q2344$+$12,
%because [Fe/Si] $>$ $-$0.2.
For these objects the ``WD low''
model is a hybrid of {\bc}=0
heating rates and {\em maximal} depletion.} 
The 
``WD-high'' model combines dust composition (2) 
with maximal
depletion: this model 
produces the highest heating rates.
For completeness we combine the Bakes \& Tielens (1994)
model of carbonaceous grains with the assumption
of maximal depletion: this model, referred to as
the ``BT'' model,  produces
heating rates between the predictions of the other two models,
and is one of the ``Gal'' models considered by WPG.
The properties of the models are summarized in Table 1.

The background mean intensities at redshift $z$, $J_{\nu}(z)$,
were computed with the software package CUBA written 
and made available to us by 
Haardt \& Madau (2003).
These authors give a detailed
description of how $J_{\nu}(z)$ 
is calculated from the luminosity function and spectra
of QSOs and how the radiation is reprocessed by the intervening gas
(Haardt \& Madau 1996). Because grain photoelectric heating
depends on $J_{\nu}$ at photon energies, $h{\nu}$ $<$ $h{\nu}_{H}$,   
where galaxies dominate QSOs,
the galaxy contribution to the background radiation is crucial.
For this reason
we considered
backgrounds that include the integrated contribution
of galaxies. We used models in which 
the radiation escaping from the galaxies is attenuated by
dust because this
is the radiation
that contributes
to the background intensity incident on a given {\DLA}. In
this case the inferred luminosity  per unit
bandwidth per unit  comoving volume decreases
with redshift at $z$ $>$ 1.5, as found by
Giavalisco {\etal} (2003). 
The predicted background spectra at the redshifts of five
{\DLAs} from our sample are shown in Figure 1.
The factor of $\sim$ 60 increase in $J_{\nu}$ as $\nu$ {\em decreases}
across $\nu_{H}$ is
the most striking feature in Figure 1. This feature is missing
from backgrounds computed from QSOs alone (see Haardt \& Madau
1996) and arises from stellar radiation, which is optically
thick at the Lyman limit,
and from an assumed escape fraction from galaxies of 10$^{-1}$
at $h{\nu}$ $\ge$ $h{\nu}_{H}$ (Haardt \& Madau 2003). 
Therefore,  {\jnu} at {\hnu} $<$ {\hnuH} is typically
10$^{-20}$ {\junit}, which is 
a factor of ten higher than  
{\jnu}  produced by
QSOs alone. By comparison, the ambient FUV radiation
in the ISM is characterized by {\jnu} $\sim$ 10$^{-19}$
{\junit}.
The spectral regions that are critical for DLA heating correspond
to 6 $<$ $h{\nu}$ $<$ 13.6 eV and $h{\nu}$ $\gtrsim$ 
400 eV. Figure 1 shows that
for a given dust-to-gas ratio 
and heating efficiency the FUV heating rates should decrease
no more than 0.5 dex between $z$=1.6 and 4.3, while 
for a given {\nh} and metallicity the X-ray heating rates should
increase by no more than 0.3 dex in the same range of redshifts.

\subsection{Cooling}

Cooling rates were computed according to the prescription given
by W95 and WPG. Thus, we include cooling due to {\lya} emission,
and grain radiative recombination, which dominate at high
temperatures, $T$ $>$ 3000 K. At lower temperatures, 
excitation of fine-structure states of abundant ions
such as C$^{+}$, O$^{0}$, Si$^{+}$, and Fe$^{+}$, and 
excitation of metastable states of O$^{0}$ dominate
the C II cooling rate.
In all cases, [C II] 158 {\micron} emission, which arises through
the transition between
the $^{2}P_{3/2}$ and  $^{2}P_{1/2}$ fine-structure states
in the $2{s^{2}}2p$ term of C$^{+}$,
is the
dominant coolant in the CNM (cold-neutral medium) phase,
while {\lya} emission dominates in the WNM 
(warm-neutral medium) phase. Because CMB excitation
of the $^{2}P_{3/2}$ state in C II
increases with 
redshift, the spontaneous emission rate 
of 158 {\micron} radiation per H atom, $l_{c}$, need
not equal the cooling rate in low density gas at high
redshifts.
To separate the effects of cooling from radiative
excitation we note
that {\lclos}=$n${\lamcii}$+$({\lclos})$_{CMB}$,
where $n${\lamcii} is the cooling rate due to [C II] 158
emission: the cooling rate is defined as the difference between
collisional excitation and de-excitation rates of
the $^{2}P_{3/2}$ state times 
the excitation
energy, $h{\nu}_{ul}$, of the
$^{2}P_{3/2}{\rightarrow}^{2}P_{1/2}$
transition. The quantity ({\lclos})$_{CMB}$ is the spontaneous
emission rate of

\begin{figure}[ht]
\includegraphics[height=4.2in, width=3.4in]{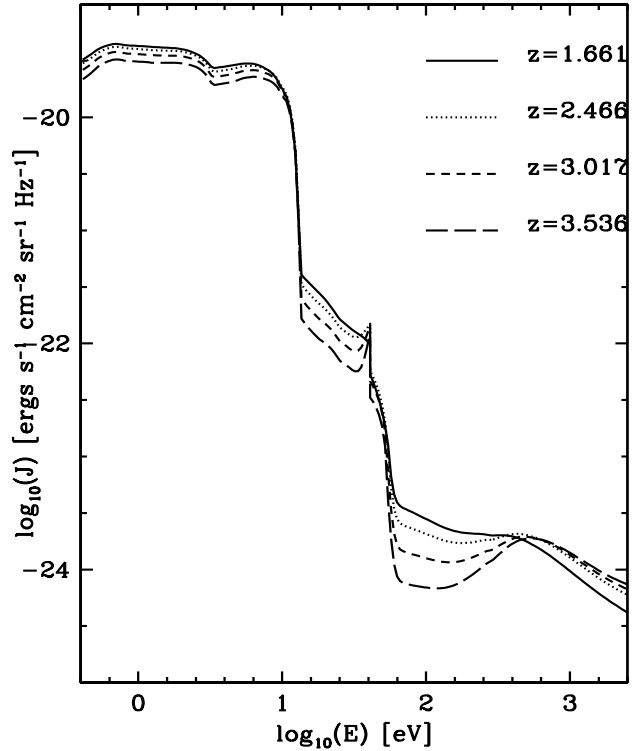}
\caption{Haardt \& Madau (2003)
background spectra for {\DLAs} with redshifts spanning
the redshift interval of our sample. Spectral features are explained
in the text} 
\label{ciistarprofile}
\end{figure}

\noindent 158 {\micron} radiation per H atom
in the limit of 
pure radiative excitation by the CMB, and
equals 2(C/H)$A_{ul}{h{\nu}_{ul}}$
exp$\biggl [-h{\nu}_{ul}/[k(1+z)T_{CMB}] \biggr ]$ where 
(C/H) is the carbon 
abundance. The quantity $A_{ul}$
is the emission coefficient for spontaneous photon decay, and
the current temperature of the CMB is $T_{CMB}$=2.728 K.

\section{HEATING BY EXTERNAL BACKGROUND RADIATION}

In this section we apply the formalism of $\S$ 2 to study
the thermal structure of {\DLAs} subject to external background
radiation.

\subsection{Background Heating of the $z$ = 1.919 Damped {\lya}
System Toward
Q2206$-$19 }

\begin{figure*}[ht]
\begin{center}
%\scalebox{0.65}[0.5]{\rotatebox{-90}{\includegraphics{f2.eps}}}
\scalebox{0.65}[0.6]{\rotatebox{-90}{\includegraphics{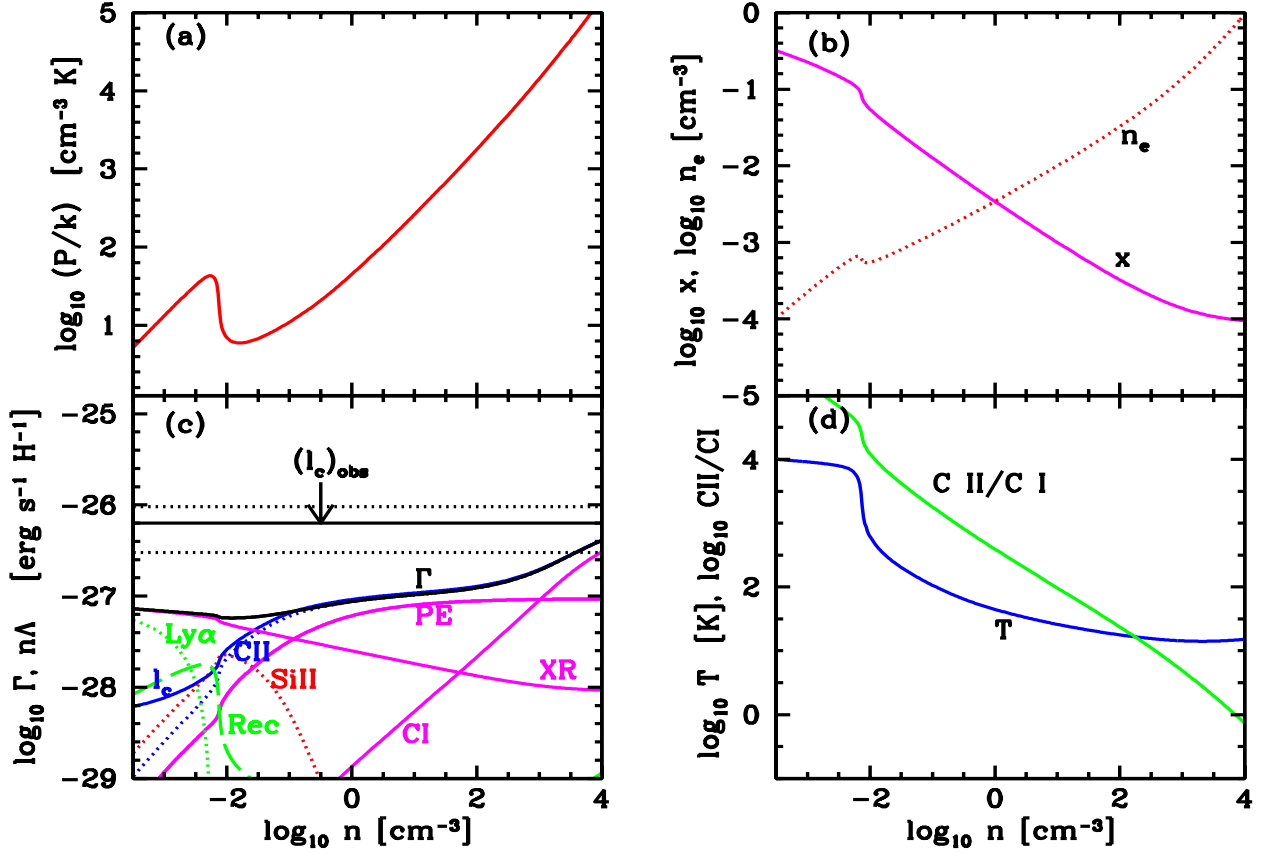}}}
\caption{Two-phase diagrams for gas in the $z$=1.919 {\DLA}
toward Q2206$-$19. The heat input is FUV and
soft X-ray  background radiation. The heating mechanisms are
grain photoelectric emission for FUV radiation and
photoionization of H, He, and abundant elements for the
X-rays. In this case we assume the ``WD-low'' model and
minimal depletion (see text). The relevant data for this
{\DLA} are in Table 2.
Panel (a) shows pressure versus density. 
Panel (b) shows fractional
ionization, $x$, and electron
density, $n_{e}$  versus density. 
Magenta curves in panel (c) show grain photoelectric heating (PE),
X-ray heating (XR), and C I photoionization heating rates versus density.
Dotted blue curve is [C II] 158 {\micron} fine-structure cooling rate, and 
green curves are {\lya} and grain recombination cooling rates.
Dotted red curve is [S II] 35 {\micron} fine-structure cooling rate.
Solid  blue curve
is [C II] 158 {\micron} spontaneous energy emission rate. Black curve ($\Gamma$) is total heating rate. Black horizontal 
solid line is log$_{10}${\lcobs} and black dotted
horizontal lines are 2-$\sigma$ errors in log$_{10}${\lcobs}
Panel (d) shows temperature and C II/C I ratio versus density. }
\label{twophase}
\end{center}
\end{figure*}

We tested the background heating hypothesis by comparing
the predicted  values of  
$l_{c}$ with observations. The observed
rate, {\lcobs} = $N$({\ciis})${A_{ul}}{h{\nu}_{ul}}$/{\nh}
where $N$({\ciis}) is the observed column density of C II
in the excited $^{2}P_{3/2}$ fine-structure state. 
The {\ciis} column densities were derived from accurate measurements of 
{\ciis}
$\lambda$ 1335.7 absorption
lines arising from the $^{2}P_{3/2}$ state. 
WPG determined {\lcobs} for 33 {\DLAs}. Most of 
the data were acquired with
the
HIRES echelle spectrograph 
(Vogt {\etal} 1994) 
on the Keck I 10 m telescope,
and in three objects with the UVES echelle spectrograph
on the VLT 8 m telescope. More recently, {\lcobs} was obtained
by Prochaska {\etal} (2003)
from measurements for 16 {\DLAs} with
the ESI echelle spectrograph imager (Sheinis {\etal} 2002) and
with UVES for another system (Pettini {\etal} 2002). Thus, the total
sample consists of {\lcobs} measurements for  
some 50 {\DLAs}. In this paper we shall analyze the properties
of 45  of these objects. These include 23 {\DLAs} with
positive detections of {\ciis} absorption and 22 with
upper limits on {\ciis} absorption.

To predict {\lclos} we computed thermal equilibria
resulting from the energy balance condition,
$\Gamma$=$n{\Lambda}$. The total heating rate
$\Gamma$=$\Gamma_{XR}$$+$${\Gamma}_{d}$, where
the radiation input is from XR and FUV background
radiation alone; i.e., no internal heat sources are included.
The quantities
$n$ and $\Lambda$ are gas volume density of hydrogen
and total cooling function
respectively. Thermal equilibrium constrains
the temperature, pressure, cooling rates, etc.
to be unique functions of density. It also results in {\lclos} as a 
unique function
of density, i.e., {\lcn}.

In order to interpret the results for our full sample of 45 {\DLAs}
the reader is 
referred to Figure 2 for the 
phase diagrams of
the $z$ = 1.919 {\DLA} toward
Q2206$-$19, one of our sample {\DLAs}, which is hereafter
referred to as the Q2206A {\DLA}. 
The figure shows equilibrium curves for (a) pressure, (b) electron
fraction and electron density, (c) cooling rates, heating rates,
and 
{\lclos}, and
(d) temperature and C II/C I
column-density ratios (hereafter C II/C I
denotes the ratio  $N$(C II)/$N$(C I)) as functions
of hydrogen density. The equilibria are computed for
the ``WD-low'' model. 
The curves show the classical  two-phase equilibria
(see Field, Goldsmith, \& Habing 1969;
W95) exhibited by neutral gas layers
heated by the grain photoelectric mechanism and
ionized by soft X-rays (W95, WPG).
In that case the equilibria admit the presence of a thermally
stable WNM at low densities and high temperatures
separated from a thermally stable
CNM at high densities and low temperatures by
a thermally unstable intermediate-density region in which
$\partial(logP)$/$\partial (logn)$ $<$ 0.
\footnote{The thermally unstable region occurs at lower
densities than in the ISM of the Galaxy because the FUV
heating  rate produced by external background radiation
incident on gas with a low dust-to-gas
ratio is about 100 times lower than in the
ISM where the heat source is internal FUV radiation
emitted by stars in the Galaxy (see WPG).}
Figure 2c 
shows  $n{\Lambda}_{CII}$ (dotted blue curve)
to increase with density. The increase is steep 
at $n$ $<$ 10$^{-2}$  cm$^{-3}$, very steep at $n$ $\approx$ 
10$^{-2}$ cm$^{-3}$,
flatter at 10$^{-2}$ $<$ $n$ $<$ 10$^{-1}$ cm$^{-3}$,
insensitive to $n$ at 10$^{-1}$ $<$ $n$ $<$ 10$^{+2}$ cm$^{-3}$,
and steeper again at $n$ $>$ 10$^{+2}$ cm$^{-3}$.

To understand this behavior we note
that the cooling rate is  dominated
by processes other than 158 {\micron} emission
at $n$ $<$ 10$^{-1}$ cm$^{-3}$. Consequently,
$n${\lamcii} does not track the total heating rate, $\Gamma$,
which is flat
at the latter densities. 
The explicit form of $n${\lamcii}
under these conditions is given by
\begin{equation}
n{\Lambda_{\rm CII}} \  {\propto} \  n_{e}{\rm (C/H)}\Omega(^{2}P_{3/2},
^{2}P_{1/2})T^{-1/2}{\rm exp}(-h{\nu}_{ul}/kT) 
\label{eq:nlambdacii}
\cmma
\end{equation}

\noindent where the collision strength, $\Omega(^{2}P_{3/2},^{2}P_{1/2})$, 
is relatively insensitive to changes in $n$ and $T$
(see Blum \& Pradhan 1992). At
$n$ $<$ 10$^{-2}$ cm$^{-3}$ the increase in $n${\lamcii}
is driven primarily by the increase in electron density 
with increasing $n$ (see Figure 2b)
because (1) the Boltzmann factor, exp($-{h{\nu_{ul}}/kT}$), is constant since
$T$ $>>$ $h{\nu}_{ul}/k$= 92 K, and (2) $T$ is insensitive
to $n$ at these densities (see Figure 2d). At $n$ $\approx$ 10$^{-2}$ cm$^{-3}$
the very steep drop in $T$ marks the transition between WNM and  CNM
gas in the region of thermal instability
due to the shift from {\lya} 
cooling to fine-structure line
cooling by low ions of abundant elements.  The  sharp 
increase in $n${\lamcii} is due to the increase
in the $T^{-1/2}$ factor because at these 
temperatures the Boltzmann
factor is constant. The decrease
in $T$ also increases 
the radiative recombination coefficients, which cause the abrupt drops
seen in $x$, $n_{e}$, and C II/C I.
When
10$^{-2}$ $<$ $n$ $<$ 10$^{-1}$ cm$^{-3}$ the increase in $n${\lamcii}
is modulated by the Boltzmann factor 
which decreases as
$T$ becomes comparable to $h{\nu}_{ul}/k$.
However, 
at $n$ $>$ 10$^{-1}$ cm$^{-3}$
the dependence of $n${\lamcii} on $n$
is dictated by the density dependence of the
total heating rate, $\Gamma(n)$, since $n${\lamcii}
dominates the cooling rate at these densities and
thus equals the total heating rate.
This accounts for the flat behavior of $n${\lamcii}
at 10$^{-1}$ $<$ $n$ $<$ 10$^{+2}$
cm$^{-3}$ where the grain photoelectric
heating rate dominates and is independent of density, and the steep
increase of $n${\lamcii} at $n$ $>$ 10$^{2}$ cm$^{-3}$ where the C I
photoionization heating rate per atom becomes dominant. 
Note that {\lcn} departs from $n${\lamcii} only at $n$ $<$ 10$^{-2}$
cm$^{-3}$ where radiative excitations by CMB radiation dominate
collisional excitations.

Comparison between predicted 
and observed values of 
{\lclos}
in the case of the Q2206A
{\DLA}
%(henceforth referred to
%as the Q2206$-$19B {\DLA})
shows that heating and ionization 
by background radiation cannot
account for the observed 158 {\micron} cooling rate  
unless the densities are  higher than
$n$ $>$ 10$^{3.4}$
{\cm3}  for the ``WD-low''
 model shown in Figure 2 
(specifically see Figure 2c). But
{\em in the case of FUV background heating alone},
densities this large
are ruled out
by the observational constraint, C II/C I $>$ 2{$\times$}10$^{4}$. 
The connection between $n$ and C II/C I follows
from the condition of 
photoionization equilibrium, which implies 
\begin{equation}
{{\rm C \ II}\over {\rm C \ I}} 
\propto
{G_{0} \over {\alpha}({\rm C I})xn} 
\cmma
\label{eq:CIIToCI}
\end{equation}

\noindent where $\alpha({\rm C I})$
is the C I radiative recombination coefficient. 
In thermal equilibrium C II/C I
and $x$  are functions of $n$, which
are plotted in Figure 2. The low value of $G_{0}$ (=0.14)
and the large lower limit on C II/C I combine to set an
upper limit on density, 
$n$ $<$ 10$^{-2}$ cm$^{-3}$ for FUV background heating
alone. From the
{\lclos}($n$) curve in Figure 2c we see this corresponds to  
{\lclos} $<$  10$^{-27.8}$
ergs s$^{-1}$ H$^{-1}$ which should be 
compared to {\lcobs} $>$ 10$^{-26.5}$
ergs s$^{-1}$ H$^{-1}$; i.e., the 2$-$$\sigma$ lower limit denoted
by the lower dotted horizontal line. {\em This demonstrates
that background heating alone cannot account for the 
value of {\lcobs}}.

Although this discrepancy
was derived for the ``WD-low'' grain model, it is quite
general as it is insensitive to grain composition.
To see this 
we compare {\lcn} with {\lcobs} for all 45 {\DLAs} in Table 2.
In each {\DLA}  we compare the predictions of the 
``WD-low'' model with the ``WD-high'' and ``BT''
models. Before discussing the full sample we 
focus first on the Q2206A {\DLA}. 
Figure 3a (first column and second row from the top)
shows the {\lcn} predicted by each
model is more than a factor of 10 below the lower
limit on {\lcobs} at densities, $n$$<$10$^{-2}$ cm$^{-3}$,
permitted by the C II/C I constraint. The discrepancies
between {\lcn} and {\lcobs} differ slightly for each model
for reasons that are independent of grain properties.
First, notice that at $n$
$<$10$^{-2}$ cm$^{-3}$ the {\lcn} predicted by the
``WD-high'' and ``BT'' models are degenerate and higher than
predicted by the ``WD-low'' model. At such low
densities the gas is a WNM where {\lcn} $\propto$ $n_{e}$(C/H).
Because there are no unsaturated
transitions of  C II we assume 
C traces Fe and as a result we let
[C/H]=[Si/H]$+$[Fe/Si]$_{int}$ 
to determine the carbon abundance
(see $\S$ 5.1
in WPG).
From Table 1 we see that the carbon abundance, [C/H],
is lower in the
``WD-low'' model than in the
``WD-high'' or ``BT'' models because [Fe/Si]$_{int}$=$-$0.2
for the minimal depletion assumed in the ``WD-low''
model and [Fe/Si]$_{int}$=0 for the maximal depletion
assumed in the ``WD-high'' and ``BT'' models.
Therefore,  
{\lcn} is predicted to be lower in
the ``WD-low'' model. Second, the  C II/C I constraints
on {\lcn} will differ because the observational limits
on C II/C I will be lower by 0.2 dex in the ``WD-low''
model. This is because $N$(C II) will be 0.2 dex lower while
$N$(C I) will be the same, as it is inferred directly from
the strength of 
C I $\lambda$ 1656.9 absorption. The predicted  dependence
of C II/C I on $n$ will also differ in each model because
the $T(n)$ and
$n_{e}(n)$ relations differ. The important point is that the 
C II/C I
limits restrict gas heated by background radiation to the WNM
phase where the limits on {\lcn} depend only on the
carbon abundance and the assumption of thermal equilibrium,
both of which are 
independent
of grain properties. This is a generic trait of the full
sample to which we now turn.

\subsection{Background Heating of the Full Sample}

We repeated these procedures for the 44 remaining {\DLAs} for which
it was possible to  calculate thermal equilibria for all three
dust models.
%In 43 of these
%it was further possible to 
%constrain the C II/C I ratios by observations. 
This yielded phase diagrams
for a total of 23 {\DLAs} in which {\ciis} $\lambda$ 1335.7
was detected  and
22 with upper limits.
Comparisons between
{\lcobs} and the predicted {\lclos}($n$) curves are plotted
for the detections  in Figures 3a and 3b, and
for the upper limits in Figures 4a and 4b. 
Relevant data for all of the {\DLAs}
are given in Table 2. The coordinate name of the background 
QSO is given in column 1, the absorption redshift in column 2,
the value of log$_{10}$C II/C I in column 3, the observed
[Fe/H]  and [Si/H] abundance ratios in
columns 4 and 5,  log$_{10}${\lcobs} in column 6, and the mean intensity
due to starlight, {\jnustar}, in
column 

\begin{figure}[ht]
\includegraphics[height=3.8in, width=2.8in]{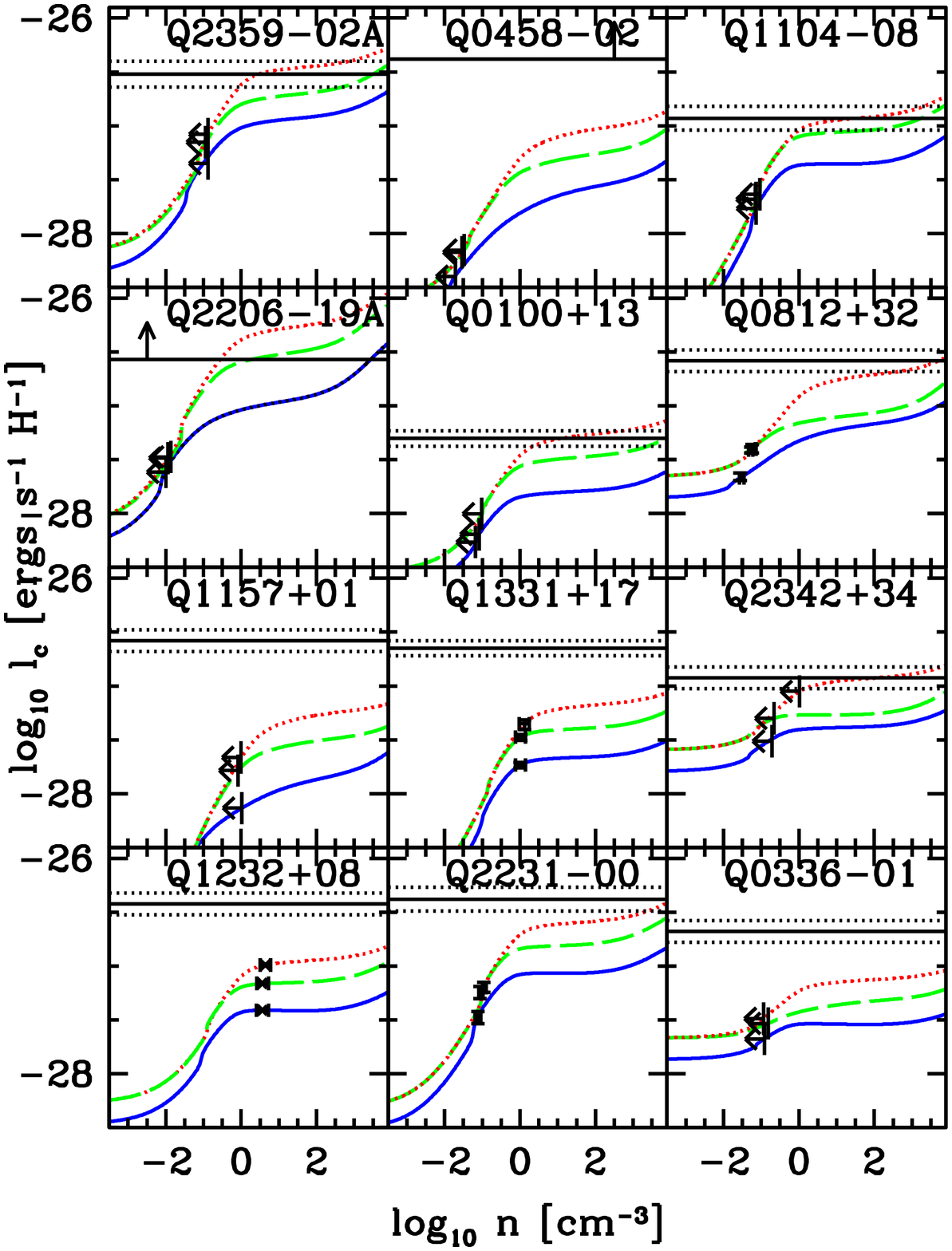}
\caption{{\lcn} curves for 12 {\DLAs} with positive
detections of {\ciis} $\lambda$1335.7 absorption.
The {\lcn} are computed from  thermal equilibria where
heat input is from background radiation alone. Solid
curves show {\lcn} for the ``WD-low'' model, dashed
curves for the ``BT'' model, and 
dotted curves for the ``WD-high'' model. 
In all cases the heating mechanisms are grain photoelectric
heating and X-ray photoionization. Data are shown as
solid horizontal lines for log$_{10}${\lcobs} and 
dotted lines for 1$-$$\sigma$ errors in 
log$_{10}${\lcobs}. The solid horizontal lines in Q2206$-$19 and Q0458$-$02
{\DLAs}
represent 2$-$$\sigma$ lower limits on log$_{10}${\lcobs}. Arrows on
{\lcn} curves are upper limits on $n$ imposed by observed lower
limit on C II/C I. These become data points and error bars  for
Q1232$+$08, Q1331$+$17, Q2231$-$00, and Q0812$+$32 
where   C II/C I is measured.}
\label{ciistarprofile}
\end{figure}

\noindent 7 (see $\S$ 5).
%Comparisons between
%{\lcobs} and the predicted {\lclos}($n$) curves are plotted
%for all 23 detections  and 9 upper limits 
%in Figures 3 and 4. 

In Figures 3a and 3b we compare the predicted {\lclos}($n$) with
{\lcobs}
for {\DLAs} with positive detections. In every case in Figure 3a
and in 9 out of 12 cases in Figure 3b the predictions
fall significantly (more than 4$\sigma_{log_{10}l_{c}}$) 
below the observations 
at densities permitted by
the  C II/C I ratios (shown as horizontal arrows directed toward
decreasing $n$).
%The results are similar to the constraints set on the
%Q2206A {\DLA}.
%namely, that 
%{\em heating by background FUV and X-ray
%radiation for the ``silicate'' and ``carbonaceous'' models is ruled
%out}.
Most of the predicted  {\lcn} curves exhibit 
behavior similar to that described for the Q2206A {\DLA}.
Specifically, the ``WD-high'' and ``BT'' models  are degenerate
at low $n$, but diverge at $n$ sufficiently large that {\lcn}
equals the grain photoelectric heating rate. At the latter
densities {\lcn} is higher for

\begin{table*}[ht] \tiny 
%\tabletypesize{\small}
\begin{center}
\caption{{\sc DLA PROPERTY}}
\begin{tabular}{lcccccc}
\tableline
\tableline
%&&\multicolumn{4}{c}{DLA Property}&\\
\cline{3-7}
    &            & log$_{10}$(C II/C I)$^{c}$&[Fe/H] & [Si/H] &  log$_{10}${\lclos}&$G_{0}$$^{f}$  \\
QSO & $z_{abs}$ &                    &      &  & erg s$^{-1}$ H$^{-1}$         \\
(1)       & (2)       & (3)        &         (4)        &  (5) &  (6)&(7)                          \\
\tableline
Q0019$-$15$^{a}$ & 3.439  &$>$3.44  & $-$1.59$\pm$0.11 &$-$1.06$\pm$0.11 &$-$26.61$\pm$0.10&6.8 \cr
Q0100$+$13$^{a}$   & 2.309  & $>$4.14 & $-$ 1.90$\pm$0.09 &$-$1.46$\pm$0.08& $-$27.30$\pm$0.07&2.4 \cr
Q0127$-$00$^{b}$   & 3.727  & $>$2.87 & $-$ 2.60$\pm$0.09 &$-$2.18$\pm$0.23& $-$27.53$\pm$0.12&1.5 (21.8) \cr
Q0133$+$04A$^{b}$   & 3.773  & $>$3.38 & $-$ 0.90$\pm$0.10 &$-$0.64$\pm$0.10& $< \ -$25.92&.... \cr
Q0133$+$04B$^{b}$   & 3.692  & $>$2.30 & $-$ 2.69$\pm$0.10 &$-$2.00$\pm$0.20& $< \ -$27.66&.... \cr
Q0149$+$33$^{a}$   & 2.141  & $>$2.80 & $-$ 1.77$\pm$0.10 &$-$1.49$\pm$0.11& $< \ -$27.24&.... \cr
Q0201$+$11$^{a}$ & 3.387  &  $>$3.22&$-$ 1.41$\pm$0.11 &$-$1.25$\pm$0.15& $-$26.67$\pm$0.14&5.4 \cr
Q0209$+$52$^{b}$   & 3.864  & $>$1.71 & $-$ 2.89$\pm$0.20 &$-$2.65$\pm$0.10& $< \ -$27.12&.... \cr
Q0255$+$00$^{a}$   & 3.915  & $>$3.23  & $-$2.05$\pm$0.10 &$-$1.78$\pm$0.05& $-$27.38$\pm$0.07&1.1 (17.3) \cr
Q0307$-$49$^{a}$   & 4.466  & $>$2.79 & $-$ 1.96$\pm$0.22 &$-$1.55$\pm$0.12& $< \ -$26.60&.... \cr
Q0336$-$01$^{a}$   & 3.062  &$>$3.86& $-$1.80$\pm$0.11&$-$1.41$\pm$0.10&$-$26.68$\pm$0.10&12.2 \cr
Q0347$-$38$^{a}$   & 3.025  & $>$3.33&$-$1.62$\pm$0.08&$-$1.17$\pm$0.03& $-$26.68$\pm$0.03&6.8 \cr
Q0405$-$44$^{b}$   & 2.595  &$>$ 4.58&$-$1.33$\pm$0.11 &$-$0.96$\pm$0.19& $ -$26.76$\pm$0.10&6.1 \cr
Q0458$-$02$^{a}$   & 2.039  &$>$ 4.60&$-$1.77$\pm$0.10 &$-$1.19$\pm$0.09& $> \ -$26.38&$>$19.1 \cr
Q0741$+$47$^{a}$   & 3.017  &$>$2.85&$-$1.93$\pm$0.10&$-$1.69$\pm$0.10&  $< \ -$27.45&....\cr
Q0747$+$27$^{b}$   & 3.900  &$>$2.44 &$-$2.53$\pm$0.10&$-$2.03$\pm$0.10&  $-$26.67$\pm$0.14&17.3\cr
Q0812$+$32$^{b}$   & 2.626  &4.07$\pm$0.10&$-$1.74$\pm$0.10&$-$0.96$\pm$0.10&$-$26.58$\pm$0.10&9.4 \cr
Q0836$+$11$^{a}$   & 2.465  &$>$3.31&$-$1.40$\pm$0.10 &$-$1.15$\pm$0.11&  $< \ -$26.98&....\cr
Q0953$+$47$^{b}$   & 4.244  &$>$1.96&$-$2.50$\pm$0.17 &$-$2.23$\pm$0.15&  $ -$26.82$\pm$0.21&17.3\cr
Q1021$+$30$^{b}$   & 2.949  &$>$2.35&$-$2.32$\pm$0.12 &$-$2.17$\pm$0.10&  $< \ -$27.31&....\cr
Q1036$-$22$^{b}$   & 2.773  &$>$3.65&$-$1.82$\pm$0.10 &$-$1.41$\pm$0.10&  $ -$27.65$\pm$0.14& (12.2)\cr
Q1104$-$18$^{a}$   & 1.661  & $>$3.87 &$-$1.48$\pm$0.10 &$-$1.04$\pm$0.10&$-$26.88$\pm$0.11&4.3 \cr
Q1108$+$22$^{a}$   & 3.608  & $>$2.70 &$-$2.12$\pm$0.10 &$-$1.80$\pm$0.11&$< \ -$27.68&.... \cr
Q1132$+$22$^{b}$   & 2.783  & $>$2.95 &$-$2.48$\pm$0.10 &$-$2.07$\pm$0.16&$< \ -$27.59&.... \cr
Q1157$+$01$^{b}$   & 1.944  & $>$3.34 &$-$1.81$\pm$0.11 &$-$1.37$\pm$0.12&$-$26.58$\pm$0.13&15.3 \cr
Q1202$-$07$^{a}$   & 4.383  &.....$^{e}$&$-$2.19$\pm$0.19&$-$1.81$\pm$0.14&$< \ -$27.06&....\cr
Q1215$+$33$^{a}$   & 1.999  & $>$3.53&$-$1.70$\pm$0.09&$-$1.48$\pm$0.07&$< \ -$27.30&....\cr
Q1223$+$17$^{a}$   & 2.466  & $>$4.07&$-$1.84$\pm$0.10&$-$1.59$\pm$0.10&$< \ -$27.02&....\cr
Q1232$+$08$^{a}$&2.337  &2.41$\pm$0.10&$-$1.72$\pm$0.13 &$-$1.22$\pm$0.15&$-$26.42$\pm$0.14&17.1\cr
Q1331$+$17$^{a}$&1.776  &2.93$\pm$0.10&$-$2.06$\pm$0.41 &$-$1.45$\pm$0.04& $-$26.65$\pm$0.07&10.8\cr
Q1337$+$11$^{b}$&2.795  &$>$2.59&$-$2.39$\pm$0.10 &$-$1.79$\pm$0.15& $< \ -$27.38&....  \cr
Q1346$-$03$^{a}$&3.736& $>$2.13 &$-$2.63$\pm$0.10&$-$2.33$\pm$0.10&$-$27.69$\pm$0.15&0.35 (15.4)\cr
Q1354$-$17$^{b}$ &2.780& $>$1.82 &$-$2.43$\pm$0.17&$-$1.88$\pm$0.16&$< \ -$26.99&....\cr
Q1506$+$52$^{b}$ &3.224& $>$2.13 &$-$2.46$\pm$0.10&$-$2.30$\pm$0.15&$< \ -$27.30&....\cr
Q1946$+$76$^{a}$ &2.884& $>$2.83 &$-$2.53$\pm$0.06&$-$2.23$\pm$0.06&$< \ -$27.33&....\cr
Q2206$-$19A$^{a}$ &1.919   & $>$4.48&$-$0.86$\pm$0.06&$-$0.42$\pm$0.07&$  -$26.20$^{+0.18}_{-0.32}$\ $^{d}$ &22.8\cr
Q2206$-$19B$^{b}$ &2.076   & $>$2.22 &$-$2.61$\pm$0.06&$-$2.31$\pm$0.07& $< \ -$26.80&.... \cr
Q2231$-$00$^{a}$ &2.066  & 3.58$\pm$0.10&$-$1.40$\pm$0.12&$-$0.88$\pm$0.10& $-$26.38$\pm$0.11&13.4 \cr
Q2237$-$06$^{a}$ &4.080   & .....$^{e}$ &$-$2.14$\pm$0.17&$-$1.87$\pm$0.11& $< \ -$27.51&.... \cr
Q2241$+$13$^{b}$   &4.282   & $>$3.07 &$-$1.90$\pm$0.11&$-$1.78$\pm$0.11& $-$27.29$\pm$0.14&.... \cr
Q2334$-$09$^{b}$ &3.057   & $>$3.29&$-$1.63$\pm$0.10&$-$1.15$\pm$0.12&$< \ -$27.04&.... \cr
Q2342$+$34$^{b}$   &2.908   &$>$3.59&$-$1.62$\pm$0.12&$-$1.19$\pm$0.11&$-$26.92$\pm$0.13&4.3 \cr
Q2344$+$12$^{b}$   &2.538   &$>$2.41&$-$1.83$\pm$0.11&$-$1.74$\pm$0.12&$< \ -$26.93&.... \cr
Q2348$-$14$^{a}$  &2.279   & $>$3.09 &$-$2.24$\pm$0.08&$-$1.92$\pm$0.08& $-$26.88$\pm$0.12&10.9 \cr
Q2359$-$02A$^{a}$  &2.095   & $>$3.52 &$-$1.66$\pm$0.10&$-$0.78$\pm$0.10& $-$26.52$\pm$0.12&8.2 \cr

\tableline
\end{tabular}
\end{center}

%\caption{TABLE 2----{\em continued}} \label{data}

\tablenotetext{a}{Data from WPG}
\tablenotetext{b}{Data presented here for the first time}
\tablenotetext{c}{Estimates derived assuming [C/H]=[Si/H], which is appropriate for the ``WD-high'' and ``BT''
model. Estimates for the ``WD-low'' models are 0.2 dex lower (see Table 1}
\tablenotetext{d}{In this case errors are 2$-$$\sigma$}
\tablenotetext{e}{Wavelength regions covering C I transitions not observed}
\tablenotetext{f}{Entries are CNM solutions for {\jnustar} ( in units
of 10$^{-19}$ {\junit}) in
{\DLAs} with positive detections. Numbers in parentheses denote WNM solutions
for cases where WNM hypothesis is plausible.}
\end{table*}

\noindent the ``WD-high'' 
model than the
``BT'' model because of the difference in
heating  efficiencies. Furthermore, {\lcn}
for both ``WD-high'' and ``BT'' models are higher than for the
``WD-low'' models because the heating
efficiencies and $\kappa$ (Table 1) are higher in
the ``WD-high'' and ``BT'' models.
However, 
there are differences with the Q2206A result.
First Figure 3b shows that in some
{\DLAs} with very low dust-to-gas ratios (e.g.
Q0953$+$47 and Q0747$+$27) {\lcn} {\em decreases} 
rather than increases with $n$
above densities corresponding to the WNM$-$CNM tran
\noindent sition
zone. This behavior stems from the low values of $\kappa$, which
suppress grain photo-electric
heating in favor of

\begin{figure}[ht]
\begin{center}
\includegraphics[height=3.8in, width=2.8in]{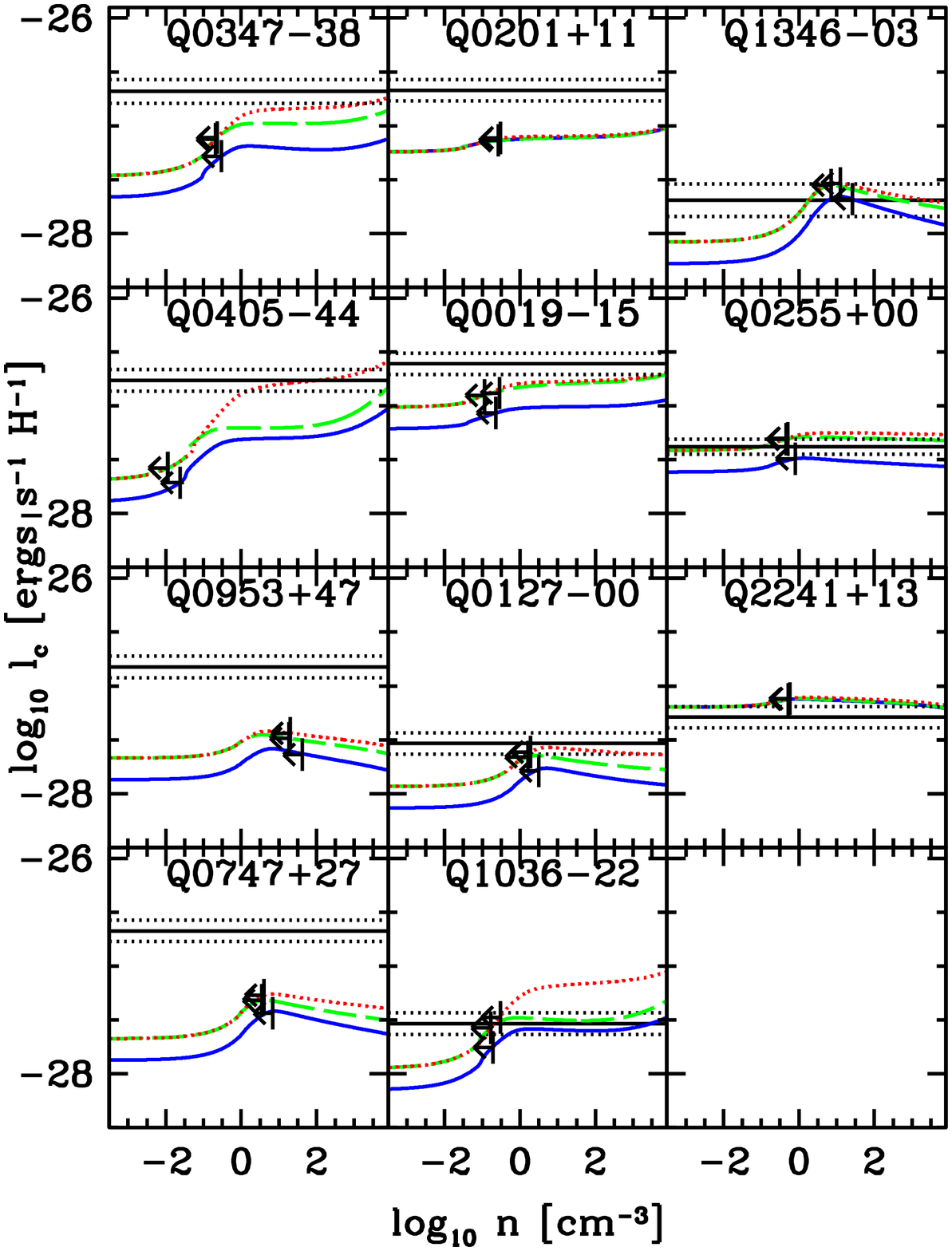}
%\caption{Same as (a) for an additional 11 {\DLAs}}
%\label{ciistarprofile}
\end{center}
\begin{center}
Fig3b
\end{center}
\end{figure}

\noindent X-ray heating, which decreases
with $n$ (see Figure 2c). Second, {\lcn} exhibits a flat
plateau at low densities in {\DLAs} with large redshifts 
%which
%is caused by the dominance of CMB radiative excitations
%over collisional excitations at sufficiently low
%values of $n$; i.e., in the low density limit
since {\lcn} approaches
({\lclos})$_{CMB}$ in the low density limit.  Because
({\lclos})$_{CMB}$ $<<$ {\lcobs},
the CMB has only marginal effect on
the [C II] 158 {\micron}  emission rate for
every {\DLA} in Figure 3a and most {\DLAs} in Figure 3b.

On the other hand, background radiation is the likely
source of 
({\lclos})$_{obs}$ for two of the 11 positive detections in
Figure 3b.
Specifically, the CMB is the dominant factor for the Q0255$+$00
and Q2241$+$13 {\DLAs} where ({\lclos})$_{CMB}$ 
$\approx$ ({\lclos})$_{obs}$. It accounts for  
{\lcobs} at all densities in all three grain models 
for Q2241$+$13
\footnote{The ``WD-low'' model for the Q2241$+$13 {\DLA}
is degenerate with the ``WD-high'' and ``BT'' models in 
Figure 3b since we assumed
maximal rather than minimal depletion. This occurs because
[Fe/Si]$_{gas}$ $>$ $-$ 0.2 in this and in the Q0201$+$11,
Q1021$+$30, Q1506$+$52,
and Q2344$+$12 {\DLAs}.}
, and it accounts for
{\lcobs} in the ``WD-high'' and ``BT'' models
for Q0255$+$00. But since the ``WD-low'' model predicts
{\lclos}($n$) to be 2.6{$\sigma_{log_{10}l_{c}}$} below
{\lcobs} at low $n$, 
a heat source in addition to background radiation  
may be required for the latter {\DLA}. 
The CMB is unlikely to explain {\lcobs} for any of the remaining
{\DLAs} with positive detections of {\ciis} absorption.
Therefore, with one exception we find that
log$_{10}${\lcobs}$-$log$_{10}${\lcn} $\ge$
3{$\sigma_{log_{10}{l_{c}}}$} at densities permitted by
the C II/C I constraint when {\lcn} is calculated
under the assumption that external background radiation is the sole
source of heating. 
The exception is the
Q1346$-$03 {\DLA} (see Figure 3b) in which
heating by
X-ray background radiation can account for {\lcobs} 
when 1 $<$ $n$ $<$ 25
cm$^{-3}$.  {\em  Therefore,
we conclude that background radiation alone
cannot explain the inferred 158 {\micron} emission 
in 20 out of the 23 positive detections in Figures 3a and 3b.
In each of the 20 cases an internal source of heating is required.}

Figures 4a and 4b  show  {\lcn} curves predicted
for the 22 {\DLAs} with
upper limits on {\lcobs}.
As for the {\DLAs} with positive detections,
{\lcobs}
exceeds {\lcn} in most cases.
But there is an important
difference: since {\lcobs} in Figures 4a and 4b are upper
limits, the discrepancy with the model predictions need
not require the presence of 
internal heat
sources. Thus, if the true {\lcobs} revealed by future
detections of {\ciis} absorption intersected the {\lcn} 
curves, then background heating could account
for the C II cooling rates, provided the gas density
satisfied the condition  {\lcn}={\lcobs}, which
defines the density  $n$=$n_{BKD}$.
Because
$n_{BKD}$ lies in the WNM phase for every {\DLA} shown
in Figure 4, the gas would be a WNM. Most of the {\DLAs} illustrated
in Figure 4 are candidates for this condition since the upper limits
on {\lcobs} are typically less than 0.2 dex above {\lcn}.
But, internal heating would be required 
if the true {\lcobs} 
remain 
above the {\lcn} curves for all $n$.
This is clearly possible for
the {\DLAs} toward Q1223$+$17, Q2348$-$14, Q2206$-$19A, and
Q0209$+$52 where the upper limits on {\lcobs} are typically more than
0.5 dex
above the peaks in the {\lcn} curves. In $\S$ 5.2 we show that internal
heating may also be present even when {\lcobs}
dips below the peaks of the  {\lcn} curves,  
which we emphasize are predicted for the case of background heating alone.  

\begin{figure}[ht]
\begin{center}
\includegraphics[height=3.9in, width=2.9in]{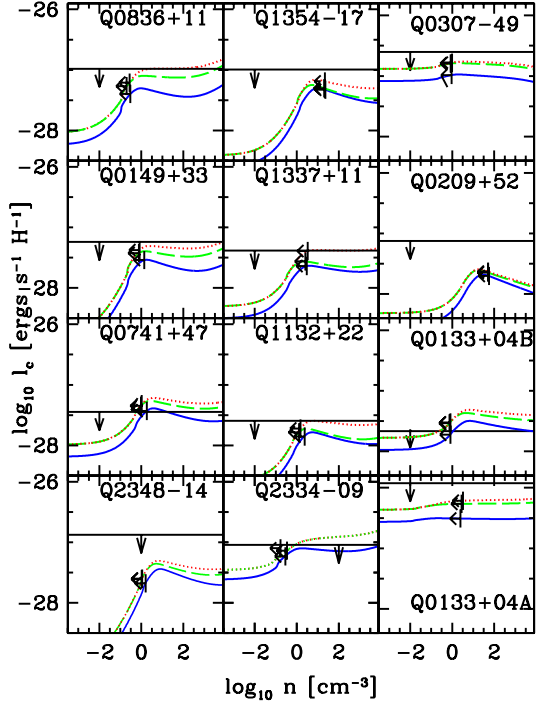}
\caption{{\lcn} curves for 12 {\DLAs} with upper limits
on {\ciis} $\lambda$ 1335.7 absorption. The {\lcn} curves
are computed as in Figures 3a and 3b.
Resulting 2$-$$\sigma$ upper limits on log$_{10}${\lcobs}
are solid horizontal lines with downward pointing arrows.}
\label{lcn1}
\end{center}
\end{figure}

\begin{figure}[ht]
\begin{center}
\includegraphics[height=3.9in, width=2.9in]{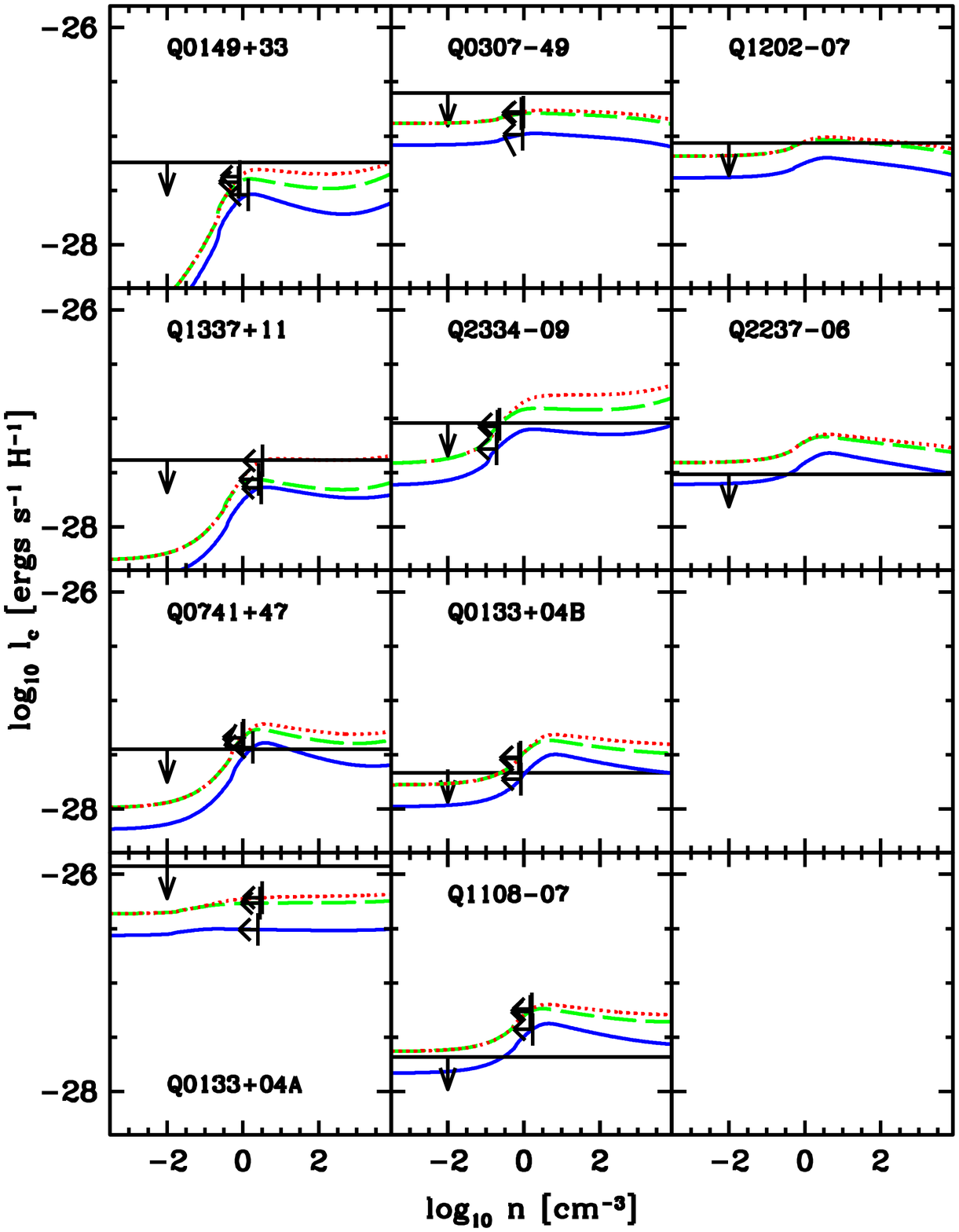}
%\caption{Same as Figure 4a for an additional 10 {\DLAs}}
%\label{ciistarprofile}
\end{center}
\begin{center}
Fig4b
\end{center}
\end{figure}

To summarize, the evidence discussed so far
suggests the following: Most {\DLAs} with positive det
\noindent ections of 
{\ciis} absorption require internal heat sources, and  
we are able to rule out the hypothesis that
damped {\lya} systems  with upper limits on {\ciis} absorption  
consist of CNM gas heated by background radiation alone. 
In $\S$ 4
we discuss independent evidence
for internal heating by starlight for one {\DLA} in 
which  
{\ciis} absorption is detected. In $\S$ 5 we consider
this possibility for all the {\DLAs} in Table 2, including those
with upper limits on {\ciis} absorption.

\section{INDEPENDENT  EVIDENCE FOR HEATING BY STARLIGHT IN A DAMPED {\lya}
SYSTEM}

In $\S$ 3.2 we concluded that heating by external background radiation
is insufficient to account for the [C II] 158 {\micron}
emission rates inferred for most  {\DLAs} with detected
{\ciis} $\lambda$ 1335.7 absorption.
Rather, an internal heat source is required. WPG and WGP
suggested grain photoelectric heating by FUV
radiation from  stars associated with the absorbing gas
as the heat source. They neglected background
radiation because the computed radiation
fields available to them included the contribution by
QSOs alone (e.g. Haardt \& Madau 1996). These fields
have a factor
of 10 lower intensity than the galaxy contribution
at {\hnu} $<$ {\hnuH} 
and, therefore, make an insignificant contribution to
the required heating rate.

Evidence supporting grain
photoelectric heating was presented in WGP. 
In particular the statistically significant correlation
between {\lcobs} and dust-to-gas ratio, $\kappa$,
argues against mechanical heating mechanisms
such as turbulent dissipation (e.g. Wolfire {\etal}
2003) and is instead strong evidence
for grain photoelectric heating, 
provided {\lcobs} represents the heating rate.
Accurate upper limits on Si II$^{*}$ $\lambda$ 1264.7 absorption
established by Howk {\etal} (2004a,b) for three {\DLAs}
with detected
{\ciis} 
$\lambda$ 1335.7 absorption give independent evidence
that the {\ciis} absorption arises
in CNM gas and, as a result, {\lcobs} must equal the
heating rate (see $\S$ 3.1). Therefore, the evidence accumulated so
far indicates (1) photoejection of electrons from grains as the
heating mechanism, and (2) that  background radiation does not provide
sufficient photons.
While this points toward locally
produced FUV radiation, i.e. starlight, as the source of photons, 
WGP did 
not offer direct evidence for a link between
the inferred radiation intensity and starlight {\em in a given {\DLA}}.

In this section we present evidence for such a link.
Specifically, M{$\o$}ller {\etal} (2002) detected
starlight emitted
by a galaxy associated with the Q2206A {\DLA}. Therefore,
it is possible to deduce {\jnu} directly and compare it with 
the {\jnu} inferred from \\ {\ciis} absorption at
{\hnu} = 8 eV. That is the purpose
of this section.

\subsection{{\jnustar} Inferred From {\ciis} Absorption}

Figure 5a shows the {\ciis} $\lambda$ 1335.7 absorption profile for
the Q2206A {\DLA}. This spectrum was acquired with UVES,
the echelle spectrograph on the VLT 8 m
telescope, by Pettini {\etal} (2002) and was
obtained by us from the UVES archive. This transition 
is clearly detected between
$v$=$-$20 and $v$=$+$30 {\kms}. At $v$ $>$$+$ 30 {\kms} the {\ciis}
transition is blended with a broad {\lya} forest feature extending
between $v$=$+$30 and $v$=$+$300 {\kms}. If one assumes that {\ciis}
$\lambda$ 1335.7 cuts off at $v$=$+$ 30 {\kms}, then 
log$_{10}$$N$({\ciis})=13.65$\pm$0.05 cm$^{-2}$. This is the
column density used to compute {\lcobs} in Figure 3.
However, this is likely to be a lower limit to the actual

\begin{figure}[ht]
\includegraphics[height=3.8in, width=2.8in]{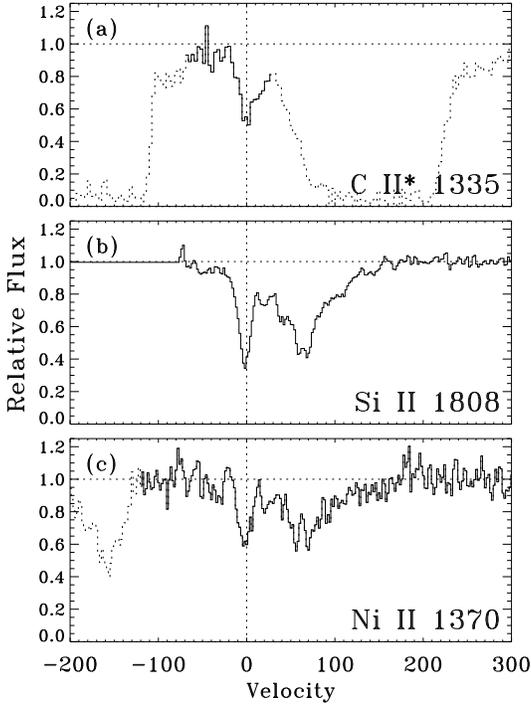}
\caption{ Panel (a) shows velocity profile of {\ciis} $\lambda$ 1335.7
absorption (solid curve) in the  Q2206A {\DLA}. 
Dotted curve at lower velocities
is C II $\lambda$ 1334.5 absorption. Dotted curve  at higher velocities
is a {\lya} forest line. Data acquired with UVES on
the VLT. Panels (b) and (c) shows Si II $\lambda$
1808.0 and Ni II $\lambda$ 1370.1 velocity profiles in the
same absorption system. Data acquired
by Prochaska \& Wolfe (1997a). Vertical line shows coincidence of the
lowest velocity component in all three transitions.}
\label{ciistarprofile}
\end{figure}

\noindent  column
density. The reasons are illustrated in Figures 5b and 5c, which
show the Si II $\lambda$ 1808.0 and Ni II $\lambda$ 1370.1 
velocity profiles (Prochaska \& Wolfe 1997a). These and all other low-ion resonance-line
profiles in this {\DLA} exhibit significant
absorption between $v$=$-$20 and $v$=$+$120
{\kms}. In fact Prochaska \& Wolfe (1997b)  used the low-ion profiles to
obtain
an absorption  velocity interval,  $\Delta v$=130 {\kms} for the Q2206A {\DLA}.
The {\ciis} profile is likely to exhibit a similar velocity
structure since the 
{\ciis} and low-ion resonance-line velocity profiles are indistinguishable
in
all but one {\DLA} for which {\ciis} $\lambda$ 1335.7 
is detected (see WGP).
When we integrate the Si II and Ni II profiles
over the entire absorption range, we find the column
densities to be 0.50 dex higher than when integrated
from $v$=$-$20 to $+$30 {\kms}. Correcting the {\ciis}
column density by this factor we find
log$_{10}$$N$({\ciis})=14.15$\pm$0.10 cm$^{-2}$. 
We take the latter value as an upper
limit on $N$({\ciis}) and  log$_{10}$$N$({\ciis}) = 13.65 cm$^{-2}$
for the lower limit. By adopting the mean of these 
for the detected value, we find
log$_{10}$$N$({\ciis})=13.97$^{+0.18}_{-0.32}$ where the
errors are the differences between the limits and the mean. 
The corresponding 158 {\micron} emission rates are
given by 
log$_{10}${\lcobs}=$-$26.20$^{+0.18}_{-0.32}$ ergs s$^{-1}$ H$^{-1}$.
We believe
the errors in log$_{10}$$N$({\ciis}) and log$_{10}${\lcobs}
correspond to 95$\%$ confidence intervals.

To obtain the 
FUV mean intensity generated by local stars 
from {\lcobs} 
we recompute the thermal equilibrium of the neutral gas. 
First we compute the grain photoelectric heating
rate, $\Gamma_{d}$, due to the total mean intensity,
{\jnu}={\jnustar}$+${\jnubkd}, where {\jnustar} and {\jnubkd} are the FUV
intensities due to local stars and background
radiation. Then we recompute the X-ray heating
rate by adding the contribution from hot gas assumed to be associated
with local stars to the X-ray background intensity
(see Wolfire {\etal} 1995 and WPG). We also
include heating by cosmic rays generated by local
stars. In both cases we let the heating rates
be proportional to the source function
for {\jnustar}; i.e., the SFR per unit area, {\ps} (see WPG). We compute
the total heating rates for a grid of mean intensities 
%we first assume the 
%gas is confined to a uniform disk in which sources of FUV radiation, i.e.,
%stars, are uniformly distributed. We use the Madau \& Pozzetti (2000)
%calibration to convert {\ps} to 
%the luminosity density
%per unit projected area, $\Sigma_{\nu}$. The latter is 
%the source function in the transfer
%equation, which we solve for {\jnu}.
\footnote{In WPG we generated the {\jnu} grid 
starting with a grid  of log$_{10}${\ps} between $-$4.0
and 0.0 {\smpykpc}. In 
this paper, we work with {\jnu} directly since it can be inferred 
from observations. We consider the SFRs later in $\S$ 6
when we connect
{\jnustar} to the rate of star formation.}
in the range
5${\times}$10$^{-21}$ $<$ {\jnustar} $<$ 5${\times}$10$^{-17}$ {\junit}.
We
use the prescription discussed in $\S$ 2 to obtain
metal abundances and dust-to-gas
ratios, 
which are used to generate cooling rates and heating rates
respectively. Thermal equilibrium models give
$P(n)$ and {\lcn} curves similar to those
shown in Figure 2. To obtain {\jnustar} corresponding to a given
{\lcobs} we assume {\DLAs} to be two-phase media in which
CNM gas is in pressure equilibrium with WNM gas at pressure
$P_{eq}$=$(P_{min}P_{max})^{1/2}$, where $P_{min}$ and $P_{max}$ are
the pressure minimum and maximum
exhibited by a two-phase medium
in thermal equilibrium (see Figure 2a). This results in two unique solutions
for each {\lcobs}, one in which {\ciis} absorption  arises in the CNM and
the other in the WNM. In WGP we argued against the WNM hypothesis
as it gives rise to more background radiation than 
observed at the current epoch. The argument against the WNM hypothesis 
is further supported by the absence of Si II$^{*}$$\lambda$
1264.7 absorption in three {\DLAs} (see Howk {\etal} 2004a,b). Nevertheless,
we consider the WNM solutions anew since they provide  independent 
tests of the
WNM hypothesis.

Figure 6 shows the Q2206A phase diagrams corresponding to
the CNM solution. In this case
the mean intensity resulting in 
{\lcn}={\lcobs} at the CNM density, {\ncnm}, is given by
{\jnustar}=2.3{$\times$}10$^{-18}$ {\junit}, where {\ncnm} 
is the higher density root of the polynomial,
$P(n)-P_{eq}$=0. We find {\ncnm}=1.3 cm$^{-3}$ as shown
by the vertical dot-dashed line which crosses the intersections 
between {\lcn} with {\lcobs} and $P(n)$ with $P_{eq}$ (Figures 6a and 6c).  
We again adopt the ``WD-low'' model.
%because of evidence against carbonaceous dust in high
%redshift {\DLAs} (e.g. Pei, Fall, \& Bechtold 1991; Cooke {\etal}
%2004). 

Comparison between
Figures 2 and 6 reveal several differences. First, 
$P_{max}$ and $P_{min}$ 
occur at higher densities when
local radiation fields with intensities much
higher than the background are  included. These pressures occur
where  fine-structure cooling
dominates resonance line cooling and signify
the transition from WNM to CNM gas. In the CNM gas the temperature
is determined by the balance between 
[C II] 158 {\micron} cooling  and
grain photoelectric heating
in the low temperature limit.
Because the density threshold above which grain photoelectric
heating dominates
X-ray and cosmic ray heating
is proportional to {\jnustar}, the transition to CNM occurs at 
higher densities as {\jnustar} increases (compare Figures 2c and 6c).
The larger value of {\jnu} also results in larger C II/C I
ratios at a given $n$ (compare Figures 2d and 6d).  Therefore, 
the high density solutions 
ruled out for background heating alone are allowed when local
sources are included. Indeed the value of {\ncnm} is consistent with the lower
limit set on C II/C I for this {\DLA}. 
 
To make realistic predictions of  
{\jnustar} for the ``WD-

\begin{figure*}[ht]
\begin{center}
%\scalebox{0.65}[0.5]{\rotatebox{-90}{\includegraphics{f6.eps}}}
\scalebox{0.65}[0.6]{\rotatebox{-90}{\includegraphics{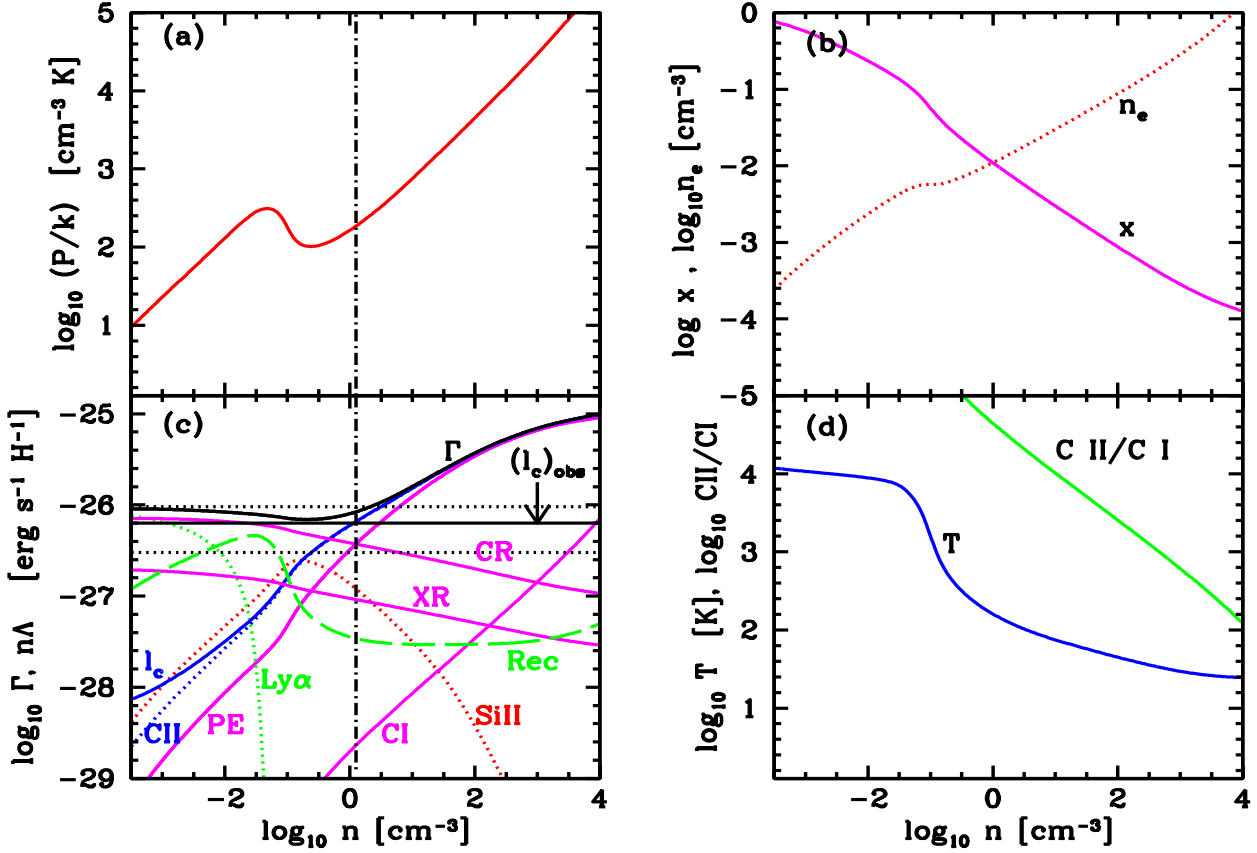}}}
\caption{Two-phase diagrams for the $z$ = 1.919  {\DLA}
toward Q2206$-$19. Same as Figure 2 except internal sources 
of heating are included. Specifically, grain photoelectric
heating by starlight with
FUV mean intensity, 
{\jnustar}=1.7{$\times$}10$^{-18}$ {\junit} is included. 
Heat inputs from locally generated soft X-rays and cosmic
rays are also included (see text). This CNM solution yields agreement
between {\lcn} and {\lcobs} at the 
CNM density, {\ncnm}=1.3 cm$^{-3}$. This is the thermally
stable density in which CNM and WNM gas are in pressure
equilibrium at pressure, $P$=$(P_{min}P_{max})^{1/2}$ 
(see text).}
\label{twophase}
\end{center}
\end{figure*}

\noindent low'' model we proceeded
as follows. First, we computed {\jnustar}  
corresponding to {\lcobs}. We repeated the calculation
for the 95 $\%$ confidence lower and upper limits
on 
{\lcobs}. We then assumed the errors in {\jnustar} to
be the differences between the limiting values of {\jnustar}
and that corresponding to the mean {\lcobs}. These errors 
%the CNM solutions 
%to be {\jnustar}= 9.2{$\times$}10$^{-19}$ and 3.6{$\times$}10$^{-18}$
%{\junit}
%indicating
%that {\jnustar}=(2.3$^{+1.3}_{-1.4}$){$\times$}10$^{-18}$ {\junit}.
%The analogous WNM solution is
%{\jnustar}=(5.1$^{+2.1}_{-2.6}$){$\times$}10$^{-17}$ {\junit}.
are due to systematic uncertainties in {\lcobs}, and are 
larger than the random statistical errors for this {\DLA}. 
Other systematic errors arise from the uncertainty in $P_{eq}$,
$\kappa$, and {\jnubkd}. To evaluate these 
we recomputed {\jnustar}  by (1) letting
$P_{eq}$ vary between $P_{min}$ and $P_{max}$, (2) varying $\kappa$
between the conditions of minimal and maximal depletion,
and (3) varying {\jnubkd} between 0.1 and 3.0 times the
Haardt-Madau backgrounds used here, which are overestimates of
the uncertainties in {\jnubkd} (Haardt \& Madau 2003).
Again
assuming the variances are determined by the differences between
these limiting values and the mean {\jnustar}  defined by 
$P_{eq}=(P_{min}P_{max})^{1/2}$, we find that 
{\jnustar}=({1.7$^{+2.7}_{-1.0}$}){$\times$}10$^{-18}$ {\junit}
for the CNM solution  and
{\jnustar}=(3.9$^{+5.9}_{-2.3}$){$\times$}10$^{-17}$ {\junit} for
the WNM solution. The latter two are the means of the minimal
and maximal depletion ``WD-low'' solutions and the errors correspond
to 95 $\%$ confidence intervals.
In every
case the {\jnustar} predicted by the WNM solutions exceed
the 95 $\%$ confidence upper limit for our CNM solution.
These estimates for the ``WD-low'' model
are also consistent with the predictions
of most ``WD-high'' and ``BT'' solutions. Figure 7 plots the 
CNM solutions of {\jnustar} versus {\jnubkd}/({\jnubkd})$_{HM}$,
where ({\jnubkd})$_{HM}$ is the Haardt-Madau background
solution adopted here. We computed solutions for
all the three grain models used in this paper.
Comparison
with the above estimate of {\jnustar} for the CNM model
shows consistency between the ``WD-high'', ``BT'',  and ``WD-low''
models with the 
possible exception of the ``BT''
model, which predicts {\jnustar} to be slightly lower than our estimate.
\footnote{Note, {\jnustar} predicted by the ``WD-high'' model is
larger than predicted by the ``WD-low'' model. This  seemingly 
contradicts the higher heating efficiency of the ``WD-high'' model,
because {\jnustar} required to produce a given heating rate decreases
with increasing heating efficiency.
However, that argument holds for a given density. Because of
the high metallicity of the Q2206A {\DLA} and the shape of the
heating function, $\Gamma_{d}(n)$, the value of {\ncnm} is lower
for the ``WD-high'' than for the ``WD-low'' model. As the
density decreases,  {\jnustar} increases in order to balance
the cooling rate, {\lcobs}.
This effect 
is more important in the Q2206A {\DLA}
than the heating heating efficiency effect,
which is why {\jnustar} is higher for the ``WD-high'' model.
Note, {\jnustar} for the ``WD-high model'' is smaller
than for the ``WD-low model'' for every other {\DLA}
in our sample.}

%With the highest heating efficiency one might have
%expected the ``extreme carbonaceous''
%model to predict lower values of {\jnustar} than the
%other ``carbonaceous'' models. Figure 7 shows instead that it
%predicts the  the highest values of {\jnustar}. This is because
%the very high 
%log$_{10}${\ps}=
%$-$1.67$\pm$0.40 {\smpykpc} for the CNM solution and
%log$_{10}${\ps}=
%$-$0.32$\pm$0.40 {\smpykpc} for the WNM solution.
%We considered silicate grains
%in all cases because of evidence against carbonaceous grains
%in high-$z$ {\DLAs}. 

Finally, it is interesting that 
these are  the highest intensities derived
with the {\ciis} technique for any
{\DLA} in our sample. This may explain
why Q2206A is the only confirmed {\DLA}
yet detected in emission at high redshift.

\subsection{{\jnu} Inferred from Direct Detection of Starlight}

M$\o$ller {\etal} (2002) obtained a STIS image of the \\ Q2206$-$19
field with a very wide $V$ filter 
in which the effective central wavelength is 5851.5 {\AA}
and FWHM=4410.3 {\AA}. The image
has
an angular resolution of 0.08 arcsec. The field shows a
diffuse galaxy with an irregular shape and a bright knot
from which {\lya} emission is detected. The identification of the
galaxy with the {\DLA} follows from the redshift of {\lya}
emission, which is 
250 {$\pm$} 50 {\kms} larger than the absorption-line redshift.
With an angular separation of $\theta_{b}$=0.99 
arcsec, the impact parameter of
the knot
with respect to the QSO sightline is given by $b$=8.4 kpc for the
WMAP (Bennett {\etal} 2003) cosmological parameters adopted here.
From their STIS photometry of the galaxy,
M$\o$ller {\etal} (2002) found $V$ = 24.69$\pm$0.07
within a circular aperture radius of 0.3 arcsec,
where $V$ is the ``$V$''  magnitude corresponding to
a filter with central wavelength $\lambda$ = 5851.5 {\AA}. 
More recently, [O III] 5007 emission
with a redshift similar to that of {\lya} emission  
was detected from the object labeled ``ps'', which is located
within $\approx$ 0.8 arcsec  of the bright knot (Warren 2004). 
%By 
%fitting an exponential to the light
%distribution centered on the knot, 
%M$\o$ller {\etal} (2002) determined the half-light
%radius to be $\theta_{1/2}$=0.5$\pm$0.1 arcsec. 
%The rather large uncertainty
%is due to the irregular morphology of the galaxy. 
%Integrating the exponential profile to $\theta$=$\theta_{b}$,
%i.e., to the QSO sightline,
%we have $V_{50}$=23.40.

To test the predictions of the {\ciis} model we
determine {\jnu} at the 
the location of the absorbing gas. 
Summing over pixels coinciding with the associated
galaxy in the
STIS image, we find

\begin{equation}
 J_{\nu} = {{(1+z)^{3}} \over {4{\pi}}}{\sum\limits_{i}}{{F^{i}_{\nu_{0}}} \over {\theta_{i}^{2}}}
\cmma
\label{eq:GammXR}
\end{equation}

\noindent where we assume the physical distance between
each pixel and the absorbing gas equals its
projection on the sky.
Here $F_{\nu_{0}}^{i}$ is the flux density in the
$i^{th}$ pixel, which is separated from the brightness
centroid of the QSO
by the angle $\theta_{i}$ and detected at frequency
${\nu_{0}}$=${\nu}(1+z)^{-1}$, where the rest wavelength
corresponding to $\nu$ is $\lambda$ = 2000 {\AA}.
In the presence of dust, the sum in equation (6)
would contain the extinction correction,
exp[${\tau^{i}_{\parallel}}$
$-$${\tau^{i}_{\perp}}$], where {$\tau^{i}_{\parallel}$} 
and {$\tau^{i}_{\perp}$}
are the 2000 {\AA} dust optical depths along the line of sight to
pixel $i$ and between pixel $i$ and the absorbing gas in the
plane of the galaxy. Independent estimates of $\tau^{i}_{\parallel}$
indicate that it varies with $\theta_{i}$. The depletion pattern implied
by the ratio [Fe/Si]$_{gas}$ suggests $\tau^{i}_{\parallel}$ $\approx$ 0.05 to
0.10 along the QSO sightline. To estimate $\tau^{i}_{\parallel}$ at the
{\lya}-emitting bright knot, we refer to observations of Lyman
Break Galaxies (LBGs). Shapley {\etal} (2003) show that $E_{B-V}$ for
LBGs to decrease with increasing {\lya} equivalent width,
$W$. Since $W$ = 88 {\AA}, we infer $E_{B-V}$$\approx$ 0.06,
and as a result $\tau^{i}_{\parallel}$$\approx$
0.5 at the position of the knot. However, owing to the irregular
light distribution, the main contribution to the sum
comes from pixels closer to the QSO sightline than the knot; i.e.,
from locations where $\tau^{i}_{\parallel}$ $<<$ 0.5. Since $\tau^{i}_{\perp}$
$<<$0.5  in the limit $\theta^{i}$ $\rightarrow$ 0, the dust
correction to equation (6) can be safely ignored.

The expression for {\jnu} in eq. (6) was kindly evaluated for us by
P. M$\o$ller (2004), who summed $F_{{\nu}_{0}}^{i}/{\theta_{i}^{2}}$
from an inner
radius surrounding the QSO, $\theta_{min}$, to an outer
radius, $\theta_{max}$. We found the result to be insensitive
to the choice of $\theta_{max}$ for $\theta_{max}$ $\ge$ 0.75
arcsec,
since irregularities in the outer brightness distribution such as
the bright knot 
contributes less than 10 $\%$ to the total {\jnu}. For this
reason we let $\theta_{max}$ = 1.5 arcsec. But the results
are sensitive to the value of $\theta_{min}$ as it approaches
the radius of the QSO PSF.
We are confident the result is not dominated
by PSF subtractions errors
when $\theta_{min}$=0.36 arcsec.
In this case we obtain a lower limit of 
{\jnuphot}=3.8{$\times$}10$^{-18}$ {\junit},
where we have introduced the quantity {\jnuphot}
to avoid confusion with the {\ciis}
technique. On the other
hand when $\theta_{min}$ $\le$ 0.21 arcsec, the result
is completely dominated by PSF subtraction errors. Therefore,
we obtain the ``upper limit''
{\jnuphot}=7.0{$\times$}10$^{-18}$ {\junit} when $\theta_{min}$
= 0.21 arcsec. The latter result is not a true
upper limit because 
point sources hidden within the PSF could raise {\jnuphot}
significantly. For example, if the bright knot were located within
0.1 arcsec of the QSO sightline, its contribution to {\jnuphot}
would be a factor of 10 larger than that of the entire galaxy. 
We shall ignore this improbable configuration and assume a uniform
light distribution 
at $\theta$ $<$ $\theta_{min}$.
From our solution to the transfer equation in WPG 
(see their eqs. (14) and (18)) we find comparable contributions to
{\jnuphot} from sources on either side of the  
$\theta$ $=$ $\theta_{min}$ boundary. For this reason we conclude that 
{\jnuphot}=(3.8$\rightarrow$14.0){$\times$}10$^{-18}$ {\junit}.

%The analogous sum,
%${\sum\limits_{i}}{F^{i}_{\nu_{0}}}$ results in 
%$V$=XX-YY  for the AB magnitude of the galaxy.

%Note, the   estimates for {\jnuphot} are upper 
%lmits as the physical separations exceed the projected separations
%used to derive equation 6.
%distance between pixel $i$
%nd the intersection point is greater than or equal to
%he projected distance inferred from $\theta_{i}$; i.e.,
%he distance used
%o derive equation 6. 

\subsection{Comparison Between {\ciis} and Photometric Estimates of {\jnu}  }

Let us compare the photometric estimate of {\jnu} 
with the  predictions of  the {\ciis} technique.
From $\S$ 4.1
we have {\jnucstar}
=(1.7$^{+2.7}_{-1.0}$){$\times$}10$^{-18}$ {\junit}
for the CNM model, and 
{\jnucstar}
=(38$^{+52}_{-23}$){$\times$}10$^{-18}$ {\junit} for the WNM model,
where in the present subsection
we have substituted the quantity {\jnucstar}
for {\jnustar} to avoid confusion with the photometric results (see
Table 3).
Since these intensities are evaluated at $\lambda$=1500 {\AA},
and {\jnuphot}  obtained from the photometric technique corresponds to
$\lambda$=2000 {\AA}, the spectral shape
of the FUV con

\begin{table} [ht] \small 
\begin{center}
\caption{Mean Intensities from Photometric and {\ciis} Techniques} \label{data}
\begin{tabular}{lccc}
Technique & Approximation&{\jnu}$^{stars}$ \ $^{d}$ &Notation   \\
\tableline
Photometry&$\theta_{min}$ \ $^{a}$&$<$3.8$^{b}$&{\jnuphot}  \\
&=0.36 arcsec & &  \\
Photometry&$\theta_{min}$ \ $^{a}$&$<$14.0$^{b}$&{\jnuphot}  \\
&=0.21 arcsec \ & &  \\
{\ciis} &CNM &1.7$^{+2.7}_{-1.0}$\ $^{c}$&{\jnucstar}  \\
{\ciis} &WNM &38$^{+52}_{-23}$ \ $^{c}$&{\jnucstar}  \\

\end{tabular}
\end{center}
\tablenotetext{a}{Inner radius used to obtain {\jnu} from eq. (6). 
Outer radius, $\theta_{max}$}
\tablenotetext{}{=1.5 arcsec.}
\tablenotetext{b}{95 $\%$ confidence upper limit obtained by
equating physical and} 
\tablenotetext{}{projected 
separations between galaxy
pixels and the QSO sightline}
\tablenotetext{c}{95 $\%$ confidence errors}
\tablenotetext{d}{10$^{-18}$ [{\junit}] }
\end{table}

\begin{figure}[ht]
\begin{center}
\includegraphics[height=5.2in, width=3.6in]{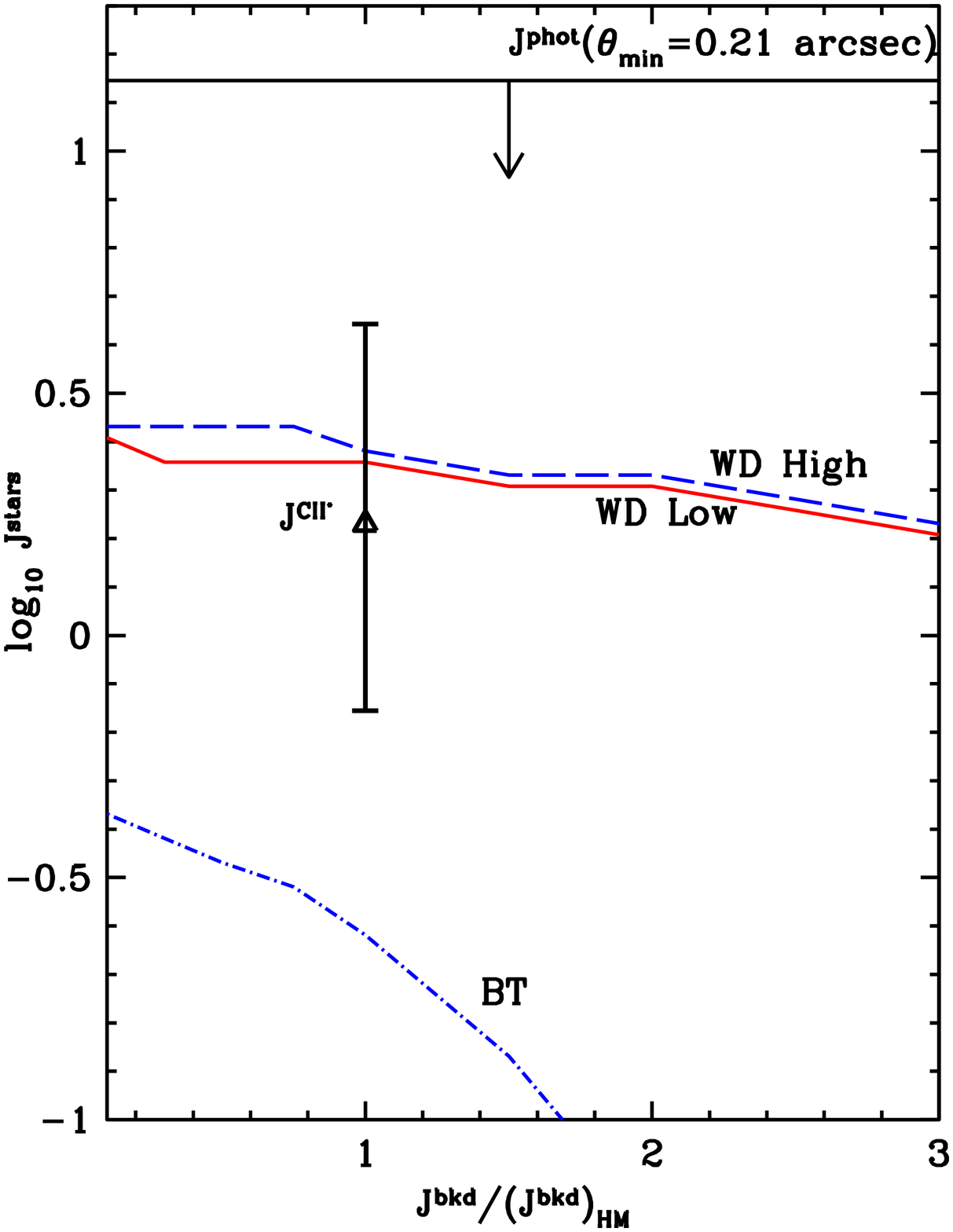}
\caption{{\jnustar} solutions for Q2206A {\DLA} versus
{\jnubkd}/({\jnubkd})$_{HM}$ where ({\jnubkd})$_{HM}$
is Haardt-Madau (2003) background radiation field used
throughout this paper and {\jnustar} is in
units of 10$^{-18}$ {\junit}. 
Curves are dot-dashed for the ``BT''
model, dashed for the ``WD-high'' model,
and solid for the ``WD-low'' model. 
Unfilled triangle labeled ``$J^{CII^{*}}$'' is our estimate
of {\jnustar} for ``WD-low'' model,
where the data point is an average of the
maximal and minimal depletion models. 
Error bars include systematic errors due to uncertainties in $\kappa$, 
$P_{eq}$, and {\jnubkd} (see text).
The ``WD-low''
solid curve was computed for minimal depletion. 
Horizontal line denoted
by $J^{phot}$($\theta_{min}$=0.21 arcsec) is upper 
limit on {\jnuphot} inferred
from photometry of associated galaxy when {$\theta_{min}$}=0.21 arcsec
(see text).}
\label{ciistarprofile}
\end{center}
\end{figure}

\noindent tinuum could be a source of additional errors.
But such errors are
negligible since the continuum is observed to be
flat  (M$\o$ller
{\etal} 2002), and theoretical models of star forming
regions predict the continuum to be flat
(Bruzual \& Charlot 2003).  
These estimates of {\jnucstar}
are compared with the estimates of {\jnuphot} 
in Table 3. 
Because the latter estimates were made by equating
the projected and physical separations between the galaxy and the absorbing
gas, the results for {\jnuphot} are upper limits.
As a result, the limits on {\jnuphot}
are consistent with the values of {\jnucstar} predicted for the
CNM, but not the WNM model.
The discrepancies with the
WNM model are larger than may be apparent because 
the error bars on {\jnucstar} correspond
to 2-$\sigma$ upper and lower limits set
by estimates of the {\ciis} column density rather than 1-$\sigma$
random errors obtained from the uncertainties in
{\nh} and to a lesser extent, $N$({\ciis}).

The
results in Table 3 are also illustrated in Figure 7,
which compares the photometric upper limit on
{\jnuphot} (horizontal line) corresponding
to $\theta_{min}$ = 0.21 arcsec with
{\jnucstar}  predicted by the CNM model (empty triangle
with vertical error bars). The true value of {\jnuphot},
({\jnuphot})$_{true}$, depends on the inclination angle, $i$, of the
galaxy to the line of sight such that
({\jnuphot})$_{true}$=cos${^2}$($i$){\jnuphot}. From Figure 7 we
see that compatibility between ({\jnuphot})$_{true}$ and
{\jnucstar} is obtained for 0{$^{\circ}$} $<$ $i$ $<$ 77{$^{\circ}$};
i.e., for a wide range of plausible inclination angles. 
In the case of the `BT' model and {\jnubkd}
$\ge$ 1.7({\jnubkd})$_{HM}$, an inclination angle,
exceeding 85$^{\circ}$ is required, which has a probability
less than 0.7 $\%$.    
Nevertheless,
the uncertainties 
are sufficiently small for us to draw an important conclusion:
we find a self-consistent model in which
[C II] 158 {\micron} cooling is balanced by the heating
rate supplied by the FUV radiation field of the galaxy
associated with this {\DLA}, provided
{\ciis} absorption arises in CNM gas.

\section{ARE ALL DAMPED {\lya} SYSTEMS INTERNALLY HEATED BY RADIATION ?}

If the Q2206A {\DLA} is heated by local starlight, 
it is plausible to ask whether all {\DLAs}
are heated in the same manner. 
This is a natural question to pose for the {\DLAs}
with positive detections of {\ciis} absorption since internal
heat sources are required in most cases (see $\S$ 3.2).
Indeed WPG assumed the presence of internal radiation fields
to deduce {\jnustar} for their sample of 33 {\DLAs}. While they
ignored the background contribution to {\jnu}, the {\lcn}
curves in Figure 3 show that background heating can be
appreciable, especially for the ``WD-high'' and ``BT'' models.
Therefore in $\S$ 5.1 we recompute 
{\jnustar} in the presence of background
radiation for {\DLAs} with positive detections.
We shall also examine the role of local radiation for
{\DLAs} with upper limits on {\ciis} absorption. 
In $\S$ 3.2
we showed that background heating alone could account for
{[C II]} 158 {\micron} emission from these objects if the
true value of {\lcobs} were sufficiently low. In $\S$ 5.2
we shall determine whether the upper limits are also consistent with
the presence of local radiation fields.

\subsection{Systems with Positive Detections}

We obtained {\jnustar} for 23 {\DLAs}
with positive detections of {\ciis} absorption (see Table 2).
In each case we followed the same procedures used in $\S$ 4.1,
in particular we added {\jnubkd}and {\jnustar} to compute the 
total mean intensity, {\jnu},
illuminating the gas, and then considered
only solutions in which {\ciis} absorption arises
in CNM gas at the thermally stable density, {\ncnm}.
Currently detected {\ciis} absorption  is unlikely
to arise in WNM gas for most {\DLAs} with 
positive detections (see  $\S$ 4).
In the case of the ``WD-low'' CNM model 
the {\jnustar} derived with the addition of 
background radiation was typically 0.1
dex less 
than the {\jnustar} derived without background radiation.
On the other hand, our calculations admitted no CNM solutions 
for {\jnustar} in the {\DLAs} toward Q2241$+$13 and Q1036$-$22:
this is not surprising in view of the significant
contribution of the high redshift CMB radiation to the {\lcn}
solutions shown in Figure 3b.  The largest
discrepancies for {\DLAs} that admitted CNM solutions were Q1346$-$03
and Q0255$+$00
for which 
{\jnustar} with background radiation was 0.55 dex and 0.15 dex less
than without background radiation. 
However, these four {\DLAs} more closely resemble most systems
with upper limits because 
{\lcobs} is less than or equal to the peak of the predicted
{\lcn} curve. In $\S$ 5.2 we show that in systems like this, 
the line of sight likely passes through WNM gas subjected
to a range of {\jnustar} including those inferred for the
CNM solutions.

Because of the 
low fraction of discrepant estimates of {\jnustar} and since
the discrepancies are confined to objects with the lowest values
determined for {\jnustar}, we conclude that  
background radiation has little effect on 
the SFRs derived by
WPG for the ``WD-low'' model. Specifically, the 
average {\jnustar}
decreased by only 0.04 dex when background radiation
was included in the models.
This quantity has cosmological significance since
the average {\jnustar} can be used to deduce the average 
SFR per unit area from which one can
deduce
the global
SFR per unit comoving volume (see WGP). 
The same
qualitative arguments hold for the ``WD-high'' and ``BT'' models.
The average {\jnustar} decreased by 0.15 dex when
background radiation was included in the ``BT''
model and by 0.3 dex for the ``WD-high''
model.

\subsection{Systems With Upper Limits}

\begin{figure*}[ht]
%\scalebox{0.65}[0.5]{\rotatebox{-90}{\includegraphics{f8.eps}}}
\scalebox{0.65}[0.6]{\rotatebox{-90}{\includegraphics{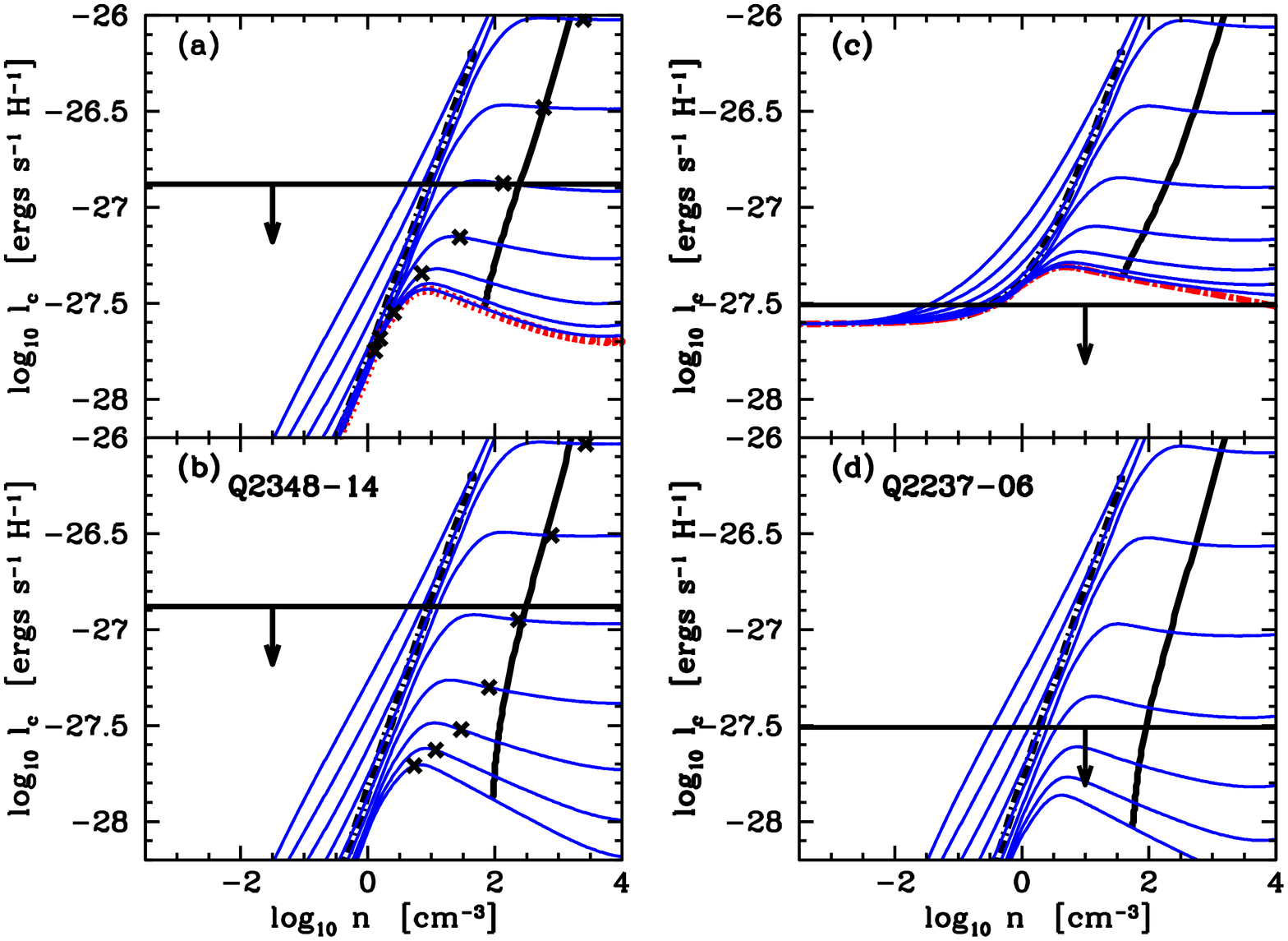}}}
\caption{Solid curves are equilibrium {\lcn} solutions
for gas subjected to
internal heating rates generated by a grid of SFRs
per unit area, log$_{10}${\ps}=$-$4.0,$-$3.5,....,0.0
{\smpykpc}. Same form of FUV, X-ray, and cosmic-ray inputs as
in Figure 6. (a) and (b) show results for 
the Q2348$-$14 {\DLA} in which background radiation is
included in (a) and excluded in (b). Lowest dotted curve in
(a) is the result for background radiation alone.
The horizontal line with arrow corresponds to 2$-${$\sigma$}
upper limit on {\lcobs}. The steeply rising  dot-dashed
and solid curves are thermally stable WNM and CNM solutions
at $n$=$n_{\rm WNM}$ and $n_{\rm CNM}$. Crosses are upper
limits on $n$ set by C II/C I constraint for each
{\ps}. (c) and (d) are analogous results for Q2237$-$06 
{\DLA}. (see $\S$ 4.2 for discussion)}
\label{twophase}
\end{figure*}

The nature of {\DLAs} with upper limits on {\ciis} absorption 
is determined by the ratio of the true value of {\lcobs} to
the peak value of {\lcn} predicted for background heating alone, 
{\lcpeak}. This point is illustrated in Figure 8, which shows
{\lcn} curves predicted by the ``WD-low'' model
for the Q2348$-$14 and
Q2237$-$06 {\DLAs}. These objects were selected
to represent the highest
and lowest values of the ratio {\lcobs}/{\lcpeak} in
Figure 4. We computed {\lcn} curves for the case 
{\jnu}={\jnubkd}. This results
in the lowest 
{\lcn} curves in Figure 8 (shown as red dotted curves), which
correspond to background heating alone. We then
let {\jnu}={\jnubkd}$+${\jnustar} to account for
the presence of local radiation and computed
{\lcn} 
for the same
grid of SFRs per unit area, $-$4.0 $<$ log$_{10}${\ps} $<$ 0.0 
{\smpykpc}, 
used to find {\jnustar}
for the Q2206A {\DLA}. This corresponds to mean 
intensities in the range
5{$\times$}10$^{-21}$ $<$ {\jnustar} $<$  
5{$\times$}10$^{-17}$ {\junit} at $\lambda$ = 1500 {\AA}
(see WPG and $\S$ 5.1). 
By comparison,
{\jnubkd} = 2.7{$\times$}10$^{-20}$
and  2.0{$\times$}10$^{-20}$ {\junit} at the redshifts
of the Q2348$-$14 and Q2237$-$06 {\DLAs}. The {\lcn} solutions
including {\jnustar} are shown as solid curves in the Figure 8.

First, we consider the results shown in Figure 8a for Q2348$-$14.
In this case the upper limit
on {\lcobs} (10$^{-26.9}$ ergs s$^{-1}$ H$^{-1}$) 
is significantly higher than 
{\lcpeak} (10$^{-27.4}$ ergs s$^{-1}$ H$^{-1}$).
If future spectroscopy shows the
true value of {\lcobs} to remain above {\lcpeak}, then
{\lcobs} could intersect WNM or CNM thermally
stable solutions corresponding to a wide range
of {\jnustar}:  the ($n$,$l_{c}$) coordinates of these
stable solutions are respectively shown as the steeply rising
dot-dashed and solid curves in Figure 8a.
The crosses in Figure 8a depict upper limits on $n$ set
by the \\ C II/C I constraints 
for each {\lcn} curve. 
Because
these density limits are less than {\ncnm}, 
they appear to rule out the CNM solutions. However, 
WGP showed the limits to be
sensitive to
local input parameters such as X-ray and cosmic-ray ionization
rates. For these reasons we make the 
conservative estimate that the CNM models are not
excluded for upper limits on $n$ exceeding $\approx$ 0.2{\ncnm}
Therefore, if {\lcobs} $>$ {\lcpeak}, 
the CNM and WNM models are in principle
equally plausible for this {\DLA}. 
But, because we have found direct evidence against the
WNM hypothesis for other {\DLAs} with similar
ratios of {\lcobs} to {\lcpeak} (see Howk {\etal} 2004 a,b)
we believe the CNM hypothesis to be more likely in these cases.

On the other hand, the CNM solutions for Q2348$-$14 are 
implausible if {\lcobs} $<$ {\lcpeak}. The presence of
background radiation causes a significant increase in
{\lcn} for low values of {\jnustar}. The result shown
in Figure 8a is that no thermally stable CNM solutions 
are possible for {\lcobs} less than the intersection
of the background heated {\lcn} and the locus of
thermally stable CNM solutions; i.e., when
{\lcobs} $<$ 10$^{-27.5}$ ergs s$^{-1}$ H$^{-1}$.
%which is only 0.1 dex below {\lcpeak}.
{\em By contrast, the WNM solutions, thermally stable or
otherwise, are consistent with the condition
{\lcobs} $<$ {\lcpeak} for the entire grid of {\jnustar}}.
The low densities, $n$ $<$ 1 cm$^{-3}$, predicted for
models with large values of {\jnustar} argue in favor of
moderate values of {\jnustar} ($<$ 10$^{-18}$ {\junit}). Otherwise
the linear dimensions of the absorbing 
gas clouds would become
implausibly large.
It is important to note that X-ray heating dominates grain
photoelectric heating in all the WNM solutions, and that
the Xray background dominates local X-ray sources for {\jnustar}
$<$ 5{$\times$}10$^{-19}$ {\junit}.

To investigate the effect of background radiation
on these conclusions,
we recomputed the {\lcn} curves with the same procedures,
but excluded all
radiation backgrounds. The results are
shown in Figure 8b. Without background heating
no lower bound on thermally stable
CNM solutions exists; i.e., the thermally stable
CNM locus can extend to arbitrarily low values of {\lcn}. 
Therefore, when the effects of background radiation are
omitted,
CNM solutions with {\lcobs} $<$ {\lcpeak}
can exist in principle.  But
when background radiation is present and
{\lcobs} $<$ {\lcpeak},
the C II/C I constraint results in upper
limits on $n$ $<$ 0.03{\ncnm}
(Figure 8a). Because this discrepancy is larger
than allowed by model uncertainties, we conclude 
that thermally stable CNM solutions are ruled
out when {\lcobs} $<$ {\lcpeak}.

Next we consider the results for Q2237$-$06 in Figure 8c.
In this case the existing upper limit on {\lcobs} is
less than {\lcpeak} and as a result all of the CNM
solutions are ruled out. Because of the large redshift,
$z$=4.080, the CMB dominates the behavior of {\lcn} for
low values of {\jnustar}. This is illustrated in Figure 8d,
which plots the  {\lcn} curves that result in the absence of
background
radiation. Comparison with Figure 8c reveals two
features caused by background radiation: a low-density asymptote
at {\lcn}=10$^{-27.6}$ ergs s$^{-1}$ H$^{-1}$ due to CMB
radiation, and an increased peak value,
{\lcpeak}=10$^{-27.3}$ ergs s$^{-1}$ H$^{-1}$, due to the
combination of CMB and X-ray background radiation. Again, Figure
8d shows that without background radiation, thermally
stable CNM solutions would be possible. 
Note, the absence of density limits 
is due to the lack of a C II/C I constraint for this {\DLA}.

Therefore, since {\lcobs}/{\lcpeak} 
$<$ 10$^{0.2}$ for most cases 
in Figures 4a and 4b, we predict the sightlines through  
most {\DLAs} with upper limits
on {\ciis} absorption to pass through WNM gas alone.
The same conclusion holds for the {\DLAs} with
positive detections for which  
{\lcobs}/{\lcpeak} are also less than 10$^{0.2}$
(i.e., the {\DLAs} toward Q0127$-$00, Q0255$+$00, Q1036$-$22, Q1346$-$03,
and Q2241$+$13). 
In the
few cases of upper limits with higher ratios of {\lcobs} to {\lcpeak},
the sightlines could encounter either WNM or CNM gas.
But the similar {\lcobs}/{\lcpeak} ratios for
{\DLAs} with positive detections,
objects for which there is independent evidence favoring
the CNM hypothesis (cf. WGP and Howk {\etal} 2004),  suggests that in
these cases the gas is
CNM.

\section{THE IMPACT OF IONIZED GAS ON  THE CII$^{*}$ ANALYSIS}

The thermal equilibria presented above were computed by assuming 
{\ciis} absorption arises in neutral gas. 
But, {\ciis} absorption   
could also arise in ionized gas. 
Since a significant fraction of the {\ciis} absorption detected in the
halo and thick gaseous disk of the Galaxy
arises in a warm ionized medium (WIM; see Lehner, Wakker, \& Savage 2004),
the same may be true in {\DLAs}. Indeed,
the presence of Al III, C IV,
and Si IV absorption in most {\DLAs} indicates that the QSO sightlines
typically pass through ionized as well as neutral gas (Wolfe \& Prochaska
2000a,b).  

However, there are several reasons why
{\ciis} absorption in {\DLAs} is unlikely  
to arise in ionized gas. 
First, from the constraint
on the optical 
depth ratio, $\tau$(Si II$^{*}$ 1264.7)/$\tau$({\ciis} 1335.7)
$<$10$^{-2.5}$,
Howk, Wolfe, \& Prochaska (2004a) deduced an upper limit
of $T$ $<$ 800 K for 
the $z$=4.22 {\DLA} toward PSS 1443$+$27. 
{\ciis} absorption cannot arise in ionized gas in this
{\DLA} because such gas typically has temperatures exceeding
10$^{4}$ K.
Because there are indications of similar limits 
in
at least two other {\DLAs}, this temperature limit may be a generic
property,  
which  would rule out ionized gas as the site of {\ciis} absorption
in most {\DLAs}. 
Second, consider the highly ionized gas, which gives rise to
C IV and Si IV absorption. In a systematic study of 32 {\DLAs},
Wolfe \& Prochaska (2000a,b) 
found statistically significant differences between the velocity structure of
the low ions and the high ions. Because
the {\ciis} velocity profiles are, with rare exceptions, 
indistinguishable from the low-ion resonance line profiles
(WPG), 
{\ciis} absorption is unlikely to originate
in highly ionized gas.
This has interesting implications. 
The presence of narrow 
C IV absorption com

\begin{figure*}[ht]
\scalebox{0.65}[0.5]{\rotatebox{-270}{\includegraphics{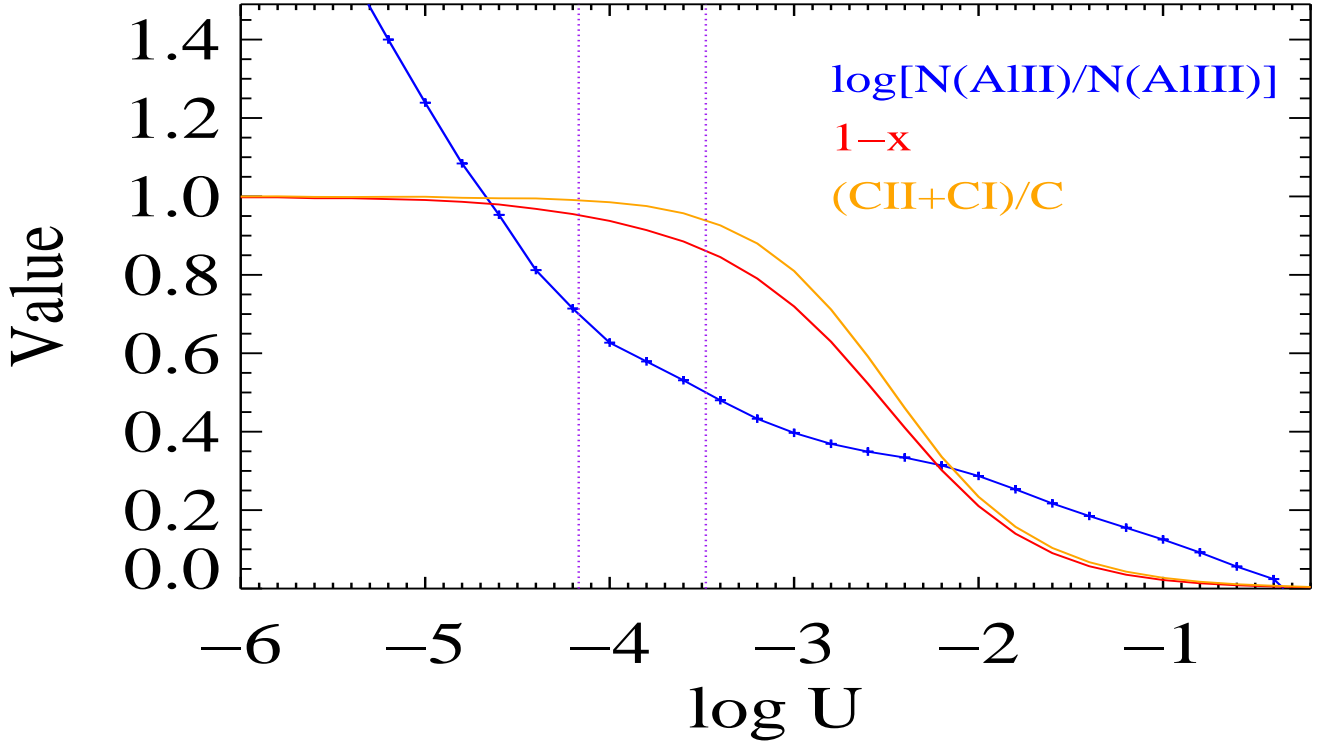}}}
\caption{Various ionization ratios versus
ionization parameter resulting from simulation of
photoionization equilibrium for 
the Q2206A {\DLA}. We let ionizing background radiation
be incident on a plane parallel slab with {\nh}=
10$^{20.5}$ cm$^{-2}$, $n$=0.1 cm$^{-3}$, and [Fe/H]
= $-$1.
Blue curve is [AlII/AlIII], red curve is linear neutral
gas fraction, 1$-x$, and orange curve is (CII$+$CI)/C.} 
\label{twophase}
\end{figure*}

\noindent ponents with FWHM $<$ 30
{\kms}  implies $T$ $<$ 2{$\times$}10$^{5}$ K,  which indicates 
CIV is photoionized in {\DLAs}.
To summarize, although hot gas may be present, comparison
between the velocity
structure of the C IV and {\ciis} absorption profiles suggests 
the two ions do not coexist in the same gas,
and   the narrow C IV line widths indicate
C IV is not directly tracing hot gas nor its
interaction with cool gas.

On the other hand, Lehner, Wakker, \& Savage (2004) present
convincing evidence that a significant fraction  of {\ciis} absorption lines
in the ISM arises in the WIM. Since the WIM of
the ISM of the Galaxy is also the site of 
Al III absorption (Savage \& Edgar 1990), the implication is that 
the gas giving rise to Al III absorption is also the site
of {\ciis} absorption. Because Al III absorption is a generic
feature in the spectra of {\DLAs}, and because of the
similarity between the centroids of the multiple velocity
components in the Al III and low-ion profiles (Wolfe \& Prochaska 2000a),
the close resemblance between the {\ciis} and low-ion
profiles suggests that {\ciis} absorption in {\DLAs} also
arises in a WIM. 
However, this is unlikely. First,
in some instances the ratio of Al III to 
low-ion column densities varies strongly with velocity (Wolfe
\& Prochaska 2000), indicating that Al III may not be an accurate
tracer of {\ciis} absorption in {\DLAs}. Second,
even if Al III does trace the low ions,
Al III need not trace WIM gas in 
{\DLAs}, since there are other means of producing
Al III than stellar photons.
At
high redshifts a primarily neutral gas can contain
significantly higher column densities of Al III than at $z$=0; i.e.,
at high $z$ the presence of Al III need not be a signature of
a WIM. This is due to the predicted 
increase with $z$ of the soft X-ray background intensity
(Haardt \& Madau 2003):
at $z$ $>$ 2 the  background intensity at 0.5 keV
(shown in Figure 1) should be more than 30 times higher than at
$z$ = 0. As a result, more radiation is available for 
photoionizing Al II (IP=18.8 eV) to Al III (IP=28.4 eV)
than at $z$=0. Prochaska {\etal} (2002) find empirical
evidence for this in 
the $z$=2.62 {\DLA} toward Q1759$+$75, where
comparison between intermediate and low-ion velocity
profiles shows significant Al III absorp
\noindent tion at
the position of the velocity component associated
with neutral gas. 

These points are illustrated in Figure 9, which shows the results
of a CLOUDY (Ferland 2001) photoionization
calculation in which we simulated 
the {\DLA} toward Q2206A with a slab
with {\nh} = 10$^{20.5}$ cm$^{-2}$, $n$=0.1 cm$^{-3}$,
and [Fe/H]=$-$1. The slab is subjected to the same
Haardt-Madau backgrounds
(2003) as shown in Figure 1. While local inputs were
neglected, these will have negligible effect on the
Al II/Al III ratio or the (CI$+$CII)/C ratio. 
The x axis is the ionization parameter, the
blue curve shows the [Al III/Al II] ratio, 
the red curve 
shows the linear neutral gas fraction, 1$-x$,
and the orange curve shows log$_{10}$(CII/C). Since we have not
measured Al II 1670 (which is undoubtedly saturated) we assume
[Al II/Fe II]=0 which is found for the few {\DLAs} in which
Al II 1670 is unsaturated. In that case the observed
ratio, [Al II/Al III] = 0.58 dex. The dotted vertical
lines show that value $\pm$ 0.1 dex. The corresponding neutral
gas fraction,
1$-$$x$, runs from 0.86 to 0.95. Therefore, 
one can accommodate a fairly large Al III column
density in a mainly neutral gas,
which produces significant amounts of C II. 
Furthermore, 
Figure 9 reveals a subtle but important point on the multi-phase
nature of the {\DLA}.  If one assumes a single phase, then the gas is
$\approx 10\%$ ionized and cannot be considered a cold neutral medium.
This result contradicts our assessment of the C\,II$^*$ profile
and we are driven to a two-phase solution (WNM+CNM) in order to accommodate
the Al\,III and C\,II$^*$ measurements.  Assuming comparable H\,I
column densities for the WNM and CNM phases, we find a solution
with $\log U \approx -4.5$ for the CNM and $\log U \approx -3$ for
the WNM, which gives an average Al II/Al III ratio consistent with
our observations.  Adopting {\jnuH} $\approx 10^{-21.5}$ for the
ionizing background radiation, 
we find the implied density values are consistent with the model presented
in Figure 6, e.g., log$_{10}${\ncnm} $\approx 0.2 {\rm cm^{-3}}$.
Note that this relatively low density for the CNM is due to the
high metallicity of the system.
In summation, the presence of Al III
does not require a WIM and in fact the observed Al II/Al III
ratio is consistent with one's expectation based on the two-phase
models derived from the C\,II$^*$ observations.

Further evidence against the ionized gas hypothesis for
{\ciis} absorption 
can be found in Figure 10, which 
plots {\lcobs} versus {\nh}
for the sample of {\DLAs} in Table 2.
%The red data points show the positive detections, 
%green are upper limits, and blue are lower limits. 
The data show no evidence for 
correlations between {\lcobs} and {\nh}: 
the Kendall tau test reveals the probability for the null
hypothesis of no correlations to be 65 $\%$.
% hile there is a clustering
%of upper limits at low column densities, this is partly a threshold
%effect in which the threshold values of detectable
%{lcobs} increases as {\nh} decreases). 
By 
contrast,
in a recent study of {\ciis} absorption in the
ISM, 
Lehner, Wakker, \& Savage (2004) find a systematic increase
in {\lcobs} with decreasing {\nh}, which they reasonably
attribute to 
an increase in ionization level in systems with
low {\nh}. To understand this we note that 
in the absence of radiative excitation, {\lcobs}
=$n{\Lambda_{\rm CII}}$(1$-x$)$^{-1}$, where
$n{\Lambda_{\rm CII}}$ is the 158 {\micron}
cooling rate {\em per particle}. In 
the low-density,
high-ionization limit applicable here, 
{\lcobs}
$\propto$ $J_{\nu_{H}}$, where
{\jnuH} is 
the radiation mean intensity at the Lyman limit.
As a result, {\lcobs} will increase with increasing
{\jnuH}, which  causes
{\nh} to decrease for a

\begin{figure}[ht]
\includegraphics[height=5.2in, width=3.7in]{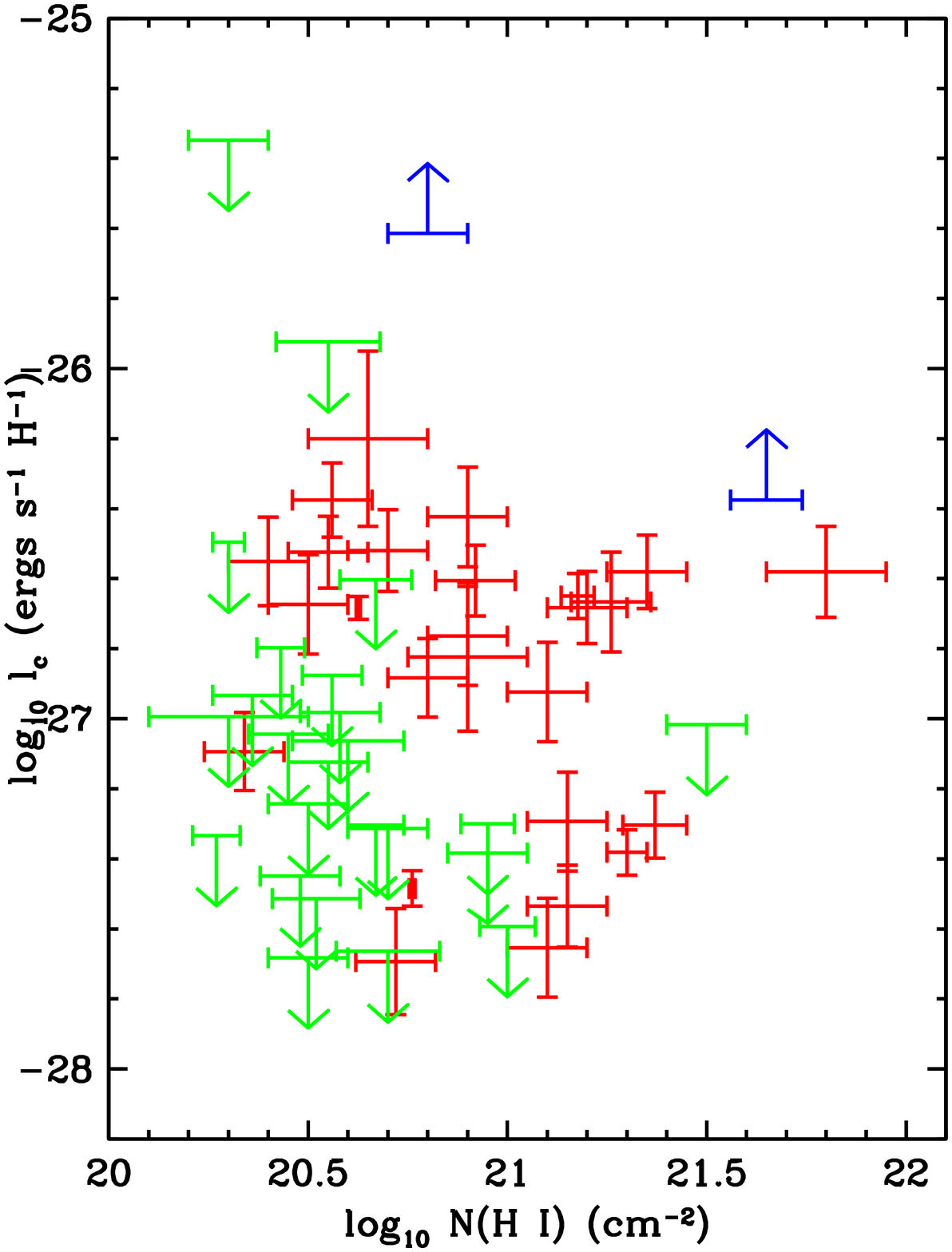}
\caption{{\lcobs} versus {\nh}. Solid red, green, and blue  data points
are positive detections, 95 $\%$ confidence upper limits, and
95 $\%$ confidence lower limits for {\DLAs} in Table 2.} 
\label{ciistarprofile}
\end{figure}

\noindent fixed total
column density.
%The elevated value of 
%{\lcobs} for the Lyman limit system
%in which {\nh}=1.3{$\times$}10$^{19}$ cm$^{-2}$ (see Figure 10)
%is evidence that this
%effect occurs at high redshift, which is consistent with
%evidence indicating the Lyman limit systems to 
%consist of ionized gas. On the other hand
By contrast, we find no convincing evidence for such an anti-correlation 
at {\nh} $\ge$ {\teno}, 
which 
is additional evidence supporting the hypothesis that
{\ciis} absorption in {\DLAs} arises in 
mostly neutral gas.

Finally, our arguments against background radiation as
the heat source source for {\DLAs} with positive detections
of {\lcobs}
depended on the upper limits on the total density, $n$,
inferred from photoionization equilibrium between C II and C I.
The C II/C I ratios and the resultant density limits were computed
for gas that was mainly neutral, and in the WNM phase for
most cases. 
While ionized gas may not be the site of
{\ciis} absorption, it may still be present and could
contribute to the total $N$(C II).
Taking the results discussed above for Q2206A, in which $x$
$<$ 0.14, and assuming $N$(C I)=0 and $N$(C II)=(C/H)$N$(H)$_{WIM}$
for the WIM (where $N$(H) is the total
column density of H), then the total C II column density,
$N$(C II)=(C/H)($N$(H)$_{WIM}$$+$$N$(H)$_{WNM}$)$\approx$(1/0.86) \\
{$\times$}(C/H){\nh}
= 1.16(C/H){\nh}.
Since the results are the same for $N$(C I), the change in the C II/C I
ratio will be at most 16$\%$. This is insufficient to alter
the conclusions
given in $\S$ 3.1.

To summarize, by contrast with the ISM, 
the evidence accumulated so far suggests that
the WIM does not give rise to significant
{\ciis} absorption in {\DLAs}. While an investigation
of the difference in physical conditions is beyond
the scope of this paper, we suggest two possibilities:
(1) an extensive WIM may
not be present in {\DLAs} owing to a low escape
fraction of ionizing radiation from 
H II regions embedded in the neutral gas.
This is consistent with the presence of Al III,
the main indicator of the  WIM in the ISM, since
Al III can arise in neutral gas in {\DLAs} due
to the increased strength of the X-ray background
at large redshifts.
%The detection of Fe III would
%be more direct evidence of a WIM (see Prochaska {\etal} 2002).
(2) If an extensive WIM does exist in {\DLAs},
{\ciis} absorption is suppressed because of the low value
of (C/H) and possibly because the
mean electron density is lower than in the
neutral ISM (see eq. 4). Lower electron
densities could result from gas pressures,
which are lower than in the CNM/WNM interface.

\section{CONCLUDING REMARKS}

Our main conclusions are: 

(1) Most {\DLAs} with positive detections
of {\ciis} $\lambda$ 1335.7 absorption 
are {\em not} heated by external background
radiation, but rather require an 
internal energy source.

(2) The internal energy
source for one such object, the Q2206A {\DLA},
appears to be  FUV radiation emitted
by stars in the associated galaxy identified by
M$\o$ller {\etal} (2002). In this case, {\ciis} absorption
cannot arise in WNM gas, but rather arises in CNM gas. 

(3) Independent arguments, in particular the
strong correlation between {\lcobs}
and dust-to-gas ratio, suggest 
{\ciis} absorption currently detected in most {\DLAs}
also arises in CNM gas that is similarly heated. 

(4) Damped {\lya} systems  with upper limits on {\ciis}
absorption also require internal energy sources if the true
values of {\lcobs} exceed the peak values of {\lcn} predicted
for heating by background radiation alone, {\lcpeak}.
While  the absorbing gas could in principle be either  CNM
or WNM, the resemblance of the ratio {\lcobs}/{\lcpeak} 
to most systems with positive detections suggests
the gas is CNM. 

(5) For most {\DLAs} with upper limits 
the true values of {\lcobs} are likely to be less than
{\lcpeak}. In this case 
the gas could be heated by background radiation
alone. But the data are also consistent with the added
presence of 
internal energy sources. In either case,
absorption is likely to  arise in 
WNM gas. The same conclusion holds for the few {\DLAs}
with positive detections and in which {\lcobs}
is also less than {\lcpeak}.

(6) In the case of radiative heating, the  radiation intensities
deduced for the positive detections and allowed by the upper limits
are about the same; i.e. 10$^{-19}$ $<$ {\jnustar} $<$ 10$^{-18}$
{\junit}. If the radiation is emitted from a uniform
distribution of stars in a disk, the range of SFRs per
unit area is given by 10$^{-3}$ $<$ {\ps} $<$ 10$^{-2}$
{\smpykpc} (see WPG). Therefore, all {\DLAs} 
may be drawn from 
a single parent population of two-phase gas layers
for which the
strength of {\ciis} absorption depends only on the
phase encountered by the line of sight through
the {\DLA}. In the case of strong {\ciis}
absorption, the gas includes a CNM with $T$ $\sim$ 100 K
and $n$ $\sim$ 10 cm$^{-3}$, and for weak absorption
the gas is WNM alone with $T$$\sim$ 8000 K and $n$ $\sim$
10$^{-1}$ cm$^{-3}$. Therefore, detectability of {\ciis}
absorption appears to depend only on the density of the gas
encountered by the line of sight.

(7) By contrast with the ISM, the evidence accumulated so far
suggests that {\ciis} absorption in
{\DLAs} does not arise in a WIM.

The arguments against background radiation as the
dominant source of heat input for {\DLAs} with positive
detections of {\ciis} absorption
are robust
as they are insensitive to the properties of the dust responsible
for grain photoelectric heating. Because of the low values of
{\jnubkd}, the large lower limits on C II/C I imply small upper
limits on $n$, typically $n$ $<$ 0.03 cm$^{-3}$
(see equation 5).
In that case the gas is restricted to the WNM phase where the
predicted 158 {\micron} emission rate, {\lcn}, is not tied
to the grain photoelectric heating rate. Instead,
{\lcn} depends on $n_{e}$, which is well determined
for the input quantities $n$ and {\jnubkd},
and the fine-structure collision strength,
which is accurately determined. We also find
that {\lcn} in the WNM is insensitive to
variations in background intensity: when we increased {\jnubkd}
by a factor of 10, {\lcn} in the WNM increased
by only 0.2 dex.
However,
to 
relate temperature
to density we assumed
thermal equilibrium, which may not be applicable
owing to the relatively long cooling times of the WNM (cf. Wolfire {\etal}
2003). But it is difficult to envisage any scenario in which
158 {\micron} emission from  
gas at such low densities depends on grain properties. 

%tracks grain photoelectric heating.
%The conclusions concerning background
%heating are also insensitive
%to the strength of the background intensity because the upper
%limits on {\lcn} are proportional to 
%${\sqrt J_{\nu}}$ in the WNM. 

By contrast, the conclusions concerning
the heat source for the Q2206A {\DLA} are sensitive to the
nature of the dust responsible for grain photoelectric heating,
because it is so metal rich.
Specifically, {\jnustar} for the ``BT''
model is between factors of 3 to 5 lower than for the other
models. The differences are related
to differences in heating efficiencies, and differences
in the predicted values of {\ncnm}.
Despite this, the agreement between our estimates of {\jnuphot}
and {\jnustar} (illustrated in Figure 7)
%which is consistent with the predictions
%of all three grain models (see Figure 7),
%factor of 10 lower than in the ``silicate'' model,
%jnd {\jnustar} vanishes for the ``extreme carbonaceous''
%jodel; both the larger values of $\kappa$ and
%jhe higher heating efficiencies of the ``carbonaceous'' models
%jincrease the input heating rates due to background radiation, thereby
%jreducing the contribution of {\jnustar} to the total {\jnu}
indicates that
FUV starlight emitted by the galaxy associated
with this DLA is incident on the grains associated with
the gas and heats the gas by the grain photoelectric mechanism.
%the reasonable agreement between 
%the photometrically determined mean intensity, {\jnuphot}, 
%and the mean intensity determined by the {\ciis} technique
%for the ``silicate'' model,
%{\jnustar},
%%and the absence of 2175 {\AA} absorption
%strongly
%suggests the Q2206A {\DLA} to be heated by FUV radiation
%emitted by the associated galaxy. 
This further argues
against mechanical heating mechanisms such as
turbulent dissipation (see discussion by  Wolfire {\etal} 2003). 
We emphasize that the agreement holds only for the CNM model, since the
WNM predictions for all three grain models indicates
{\jnustar} exceeds the 95 $\%$
confidence upper limit on {\jnuphot}.
This result is in agreement with
the conclusions reached by WPG and WGP who found that 
the bolometric background intensity predicted at $z$ = 0 if the WNM
model applies to all {\DLAs}
violated observational constraints.
We also emphasize
that the {\jnustar} predicted for this {\DLA} are the highest
in our sample, which may help to explain why Q2206A
is the only confirmed {\DLA} detected in emission
at high redshift.

Finally, we consider the  SFR of the
Q2206A {\DLA}. Summing over the same pixels
used to determine {\jnustar} in equation (6), we can estimate
the total FUV flux density of the associated galaxy. Assuming again
that
$\theta_{max}$ = 1.5 arcsec,  we find that $V$=23.0 if
$\theta_{min}$=0.36 arsec and $V$=22.9 if $\theta_{min}$ = 0.21
arcsec; i.e., the results are insensitive to the value of
$\theta_{min}$. We also find they are insensitive to the
value of $\theta_{max}$ for $\theta_{max}$ $>$ 1.0 arcsec.
Assuming $V$=23 we find that
the luminosity per unit bandwidth,
$L_{\nu}$=(2.0$\pm$0.50)$\times$10$^{29}$
ergs s$^{-1}$ Hz$^{-1}$ where we have adopted conservative
25 $\%$ errors for the photometry. As a result 
SFR = (26$\pm$6.5){\smpy}. Because we have
not corrected for the effects of extinction, this is 
a lower limit. Following the discussion in $\S$ 4.2 in
which we used the 
equivalent
width of {\lya} emission
to estimate $E_{B-V}$  we find that
SFR $<$ 50 {\smpy}.   
Comparison with LBGs, which are drawn from a
Schechter luminosity function characterized by
(SFR)$_{*}$=60 {\smpy}  and $\alpha$ = 1.6 (Steidel {\etal} 1999;
Shapley {\etal} 2003) shows the Q2206A galaxy to be
an LBG with a typical SFR. On the other hand its
irregular and extended morphology is atypical
for LBGs and may be atypical for galaxies associated
with {\DLAs}. 
For this reason it is not obvious that these results
can be extrapolated to other {\DLAs} with detected
{\ciis} absorption by simply scaling SFR with {\jnustar}.
Indeed, applying the simple uniform disk model of
WPG and WGP to Q2206A, we find {\ps} = 0.03 {\smpykpc}
rather than the observed value of {\ps} $>$ 1 {\smpykpc}.
The latter value of {\ps} is consistent with the derived
value of {\jnuphot} because of the small solid angle subtended
by the FUV continuum sources from the location
of the absorbing gas. Of course the SFR per unit
H I area would be considerably lower if the star forming
region were embedded in a larger H I envelope. 
Interestingly, at least 50 $\%$ of the remaining
{\DLAs} with detected {\ciis} absorption are predicted
to have {\jnustar} within a factor of 3 of
the {\jnustar} inferred for
the Q2206A {\DLA}. It is not clear at this time
whether or not their SFRs are within a factor
of 3 of the SFR derived for the Q2206A galaxy.
We are pursuing a program with the ACS aboard the HST to
investigate this question.

\acknowledgements

The authors wish to recognize and acknowledge the very significant
cultural
role and reverence that the summit of Mauna Kea has always had within the
indigenous Hawaiian community.  We are most fortunate to have the
opportunity to conduct observations from this mountain.
We thank Francesco Haardt and Piero Madau for
providing us with their calculations of background intensities
and for helpful discussions, and Bruce Draine and Pale
M$\o$ller for helpful discussions. 
We thank
Dan McCammon for giving us his updated tabulation of
X-ray photoionization cross-sections. 
AMW and JXP are partially supported by NSF grant AST0307824.

\newpage


\begin{thebibliography}{}

\bibitem[Bakes\ (1994)]{bakes94}
Bakes, E. L. O. \& Tielens, A. G. G. M.
1994, \apj, 427, 822

\bibitem[Bennett\ (2003)]{benn03}
Bennett, C. L., Halpern, M., Hinshaw, G., Jarosik, N., 
Kogut, A., Limin, M., Meyer, S. S., Page, L., Spergel, D. N.,
Tucker, G. S., Wollack, E., Wright, E. L., Barnes, C.,
Greason, M. R., Hill, R. S., Komatsu, E., Nolta, M. R.,
Odegard, N., Peiris, H. V., Verde, L., \& Weiland, J. L., 2003,
\apjs, 148, 1

\bibitem[blum\ (1992)]{blum92}
Blum, R. D.  \& Pradhan, A. K.
1992, \apjs, 80, 425


\bibitem[bruzual\ (2003)]{bruz03}
Bruzual, G. \& Charlot, S., 2003,
\mnras, 344, 1000

\bibitem[churchill\ (2003)]{church03}
Churchill, C. W., Vogt, S. S., \& Charlton, J. C., 2003,
\aj, 125, 98

\bibitem[cooke\ (2004)]{cooke04}
Cooke, J., Wolfe, A. M., Howk, J. C., Prochaska, J. X. 
\& Gawiser, E. 2004, in preparation.







\bibitem[draine \ (1978)]{draine78}
Draine, B. T. 1978, \apjs, 36, 595



\bibitem[Erb et al. (2003)]{erb03}
Erb, D. K., Shapley, A. E., Steidel, C. C., Pettini, M.,
Adelberger, K. L., Hunt, M. P., Moorwood, A. F. M., 
\& Cuby, J.-G.
2003, \apj, 591, 101


\bibitem[field \ (1969)]{fgh69}
Field, G. B., Goldsmith, D. W., \& Habing, H. J.  1969, 
\apj, 155, L149                        


\bibitem[Giavalisco (2004)]{giav04}
Giavalisco, M., Dickinson, M., Ferguson, H. C., Ravindranath, S.,
Kretchmer, C., Moustakas, L. A., Madau, P., Fall, S. M.,
Gardner, J. P., Livio, M., Papovich, C., Renzini, A., Spinrad, H.,
\& Riess, A.
2004, \apj, 600, L103

\bibitem[Haardt Madau\ (1996)]{hm96}
Haardt, F.  \& Madau, P. 1996, \apj, 461, 20

\bibitem[Haardt Madau\ (2003)]{hm03}
Haardt, F.  \& Madau, P. 2003, 
http://pitto.mib.infn.it/~haardt/cosmology.html

\bibitem[Howk\ (2004)]{ch03}
Howk, J. C., Wolfe, A. M., \& Prochaska, J. X.
2004a \apj, submitted (astro-ph/0404005)

\bibitem[Howk\ (2004)]{ch03}
Howk, J. C., Wolfe, A. M., Prochaska, J. X.  \&  Gawiser,
E. 2004b, in preparation

\bibitem[Junkkarinen et al. (2004)]{junk03}
Junkkarinen, V. T., cohen, R. D., Beaver, E. A., Burbidge, E. M.,
Lyons, R. W., \&  Madejski, G.  2004, \apj, in press (astro-ph/0407281)

\bibitem[Kennicutt\ (1998)]{kenn98}
Kennicutt, R. C. Jr. 1998, \araa, 36, 189

\bibitem[kulkarni (2001)]{kul01}
Kulkarni, V. P., Hill, H. M., Schneider, G., Weymann, R. J.,
Storrie-Lombardi, L. J., Rieke, M. J., 
Thompson, R. I. 2001, \& Jannuzi, B. T. 2000, \apj, 536, 36

\bibitem[kulkarni (2001)]{kul01}
Kulkarni, V. P., Hill, H. M., Schneider, G., Weymann, R. J.,
Storrie-Lombardi, L. J., Rieke, M. J., 
\& Thompson, R. I. 2001, \apj, 551, 37

\bibitem[lehner(2004)]{leh04}
Lehner, N., Wakker, B., \& Savage, B.D. 2004, preprint

\bibitem[machacek (2000)]{mach00}
Machacek, M. E., Bryan, G. L., Meiksin, A., Anninos, P.,
Thayer, D., Norman, M. L., \& Zhang, Y. 2000, \apj, 532, 118



\bibitem[Madau (1996)]{pmd96} 
Madau, P., Ferguson, H. C., Dickinson, M. E., Giavalisco, M.,
Steidel, C. C., \& Fruchter, A. 1996,
\mnras, 283, 1388 


\bibitem[Madau (2000)]{pmd20} 
Madau, P., \& Pozzetti, L. 2000,
\mnras, 312, L9

\bibitem[McCammon (2003)]{mcC03} 
McCammon, D. 2003, private communication


\bibitem[Miralda-Escud$\rm \acute e$ (1996)]{mom96}
Miralda-Escud$\rm \acute e$, J., Cen, R., Ostriker,
J. P. O, \& Rauch, M. 1996, \apj, 471, 582

\bibitem[Mo \& Miralda-Escud$\rm \acute e$ (1996)]{mom96}
Mo, H.J., \& Miralda-Escud$\rm \acute e$, J. 1996, \apj, 469, 589



\bibitem[moller (2002)]{moll 2002}
Moller, P., Warren, S. J., Fall, S. M., Fynbo, J. U.,
\& Jakobsen, P. 2002, \apj, 574,51

\bibitem[morrison \& mccammon (1836)]{mor83}
Morrison, R.,  \& McCammon, D. 1983, \apj, 270, 119

\bibitem[motta (2002)]{mot02}
Motta, V., Mediavilla, E., Munoz, J. A., Falco, E., 
Kochanek, c. S., Arribas, S., Garcia-Lorenzo, B., Oscoz, A.,
\& Serra-Ricart, M. 2002, \apj, 574, 719

\bibitem[Pei (1991)]{pfb1991}
Pei, Y. C., Fall, S. M., \& Bechtold, J.  1991, \apj, 378, 6

\bibitem[Pei (1995)]{peiF1995}
Pei, Y. C., \& Fall, S. M. 1995, \apj, 454, 69





\bibitem[Pettini (1994)]{max1994}
Pettini, M., Smith, L. J., Hunstead, R. W., \& King, D. L. 
1994, \apj, 426, 79


\bibitem[Pettini et al. (2001)]{max01}
Pettini, M., Shapley A. E., Steidel, C. C., Cuby, J.-G., Dickinson, M.,
Moorwood, A. F. M., Adelberger, K. L., \& Giavalisco, M.
2001, \apj, 554, 981

\bibitem[Pettini (2002)]{max2002}
Pettini, M., Ellison, S. L., Bergeron, J.,  \& Petitjean, P. 
2002, \aa, 391, 21


\bibitem[Prochaska and Wolfe (1997)]{pro97}
Prochaska, J. X. and Wolfe, A. M. 1997a, \apj, 474, 140 

\bibitem[Prochaska and Wolfe (1997)]{pro97}
Prochaska, J. X. and Wolfe, A. M. 1997b, \apj, 487, 73  


\bibitem[Prochaska  (1999)]{pro97}
Prochaska, J. X.  1999, \apj, 511, 71 

\bibitem[Prochaska  (2002)]{pro02}
Prochaska, J. X., Howk, J. C., O'Meara, J. M., Tytler, D.,
Wolfe, A. M., Kirkman, D., Lubin, D., \&
Suzuki, N.  2002, \apj, 571, 693 


\bibitem[Prochaska {\etal} (2003)]{pro03}
Prochaska, J. X., Castro, S., 
\& Djorgovski, S. G., 2003 \apjs, 148, 317


\bibitem[rauch \ (1998)]{rauch98}
Rauch, M. 1998, \araa, 36, 267


\bibitem[scott (2002)]{scott02}
Scott, J., Bechtold, J., Morita, M., Dobrzycki, A.,.
\& Kulkarni, V. 2002, \apj, 571, 665



\bibitem[shapley et al. (2003)]{shap03}
Shapley, A. E., Steidel, C. C., Pettini, M.,
\& Adelberger, K. L.
2003, \apj, 588, 65

\bibitem[Sheinis (2003)]{shein03}
Sheinis, A.I., Miller, J., Bigelow, B., Bolte, M.,
Epps, H., Kibrick, R., Radovan, M.,  \& Sutin, B.
2002, \pasp, 114, 851 

\bibitem[lisa (2000)]{lisa00}
Shull, J. M. \&  Van Steenberg, M. E. 1985,
\apj, 298, 268


\bibitem[Steidel (1999)]{stei99}
Steidel, C. C., Adelberger, K. L., Giavalisco, M., Dickinson,
M.,  \& Pettini, M. 1999, \apj, 519, 1

\bibitem[lisa (2000)]{lis100}
Storrie-Lombardi, L.J.,
\& Wolfe, A. M. \apj, 543, 552



\bibitem[Vogt \ (1994)]{vogt94}
Vogt, S. S., Allen, S. L., Bigelow, B. C.,
Bresee, L., Brown, B., Cantrall, T., Conrad, A.,
Couture, M., Delaney, C., Epps, H. W.,
Hilyard, D. F., Horn, E., Jern, N., Kanto, Dl, Keane, M. J.,
Kibrick, R. I., Lewis, J. W., Osborne, J. Pardeilhan, G. H.,
Pfister, T., Ricketts, T., Robinson, L. B.,
Stover, R. J., Tucker, D., Ward, J., \& Wei. M. Z.
1994, SPIE, 2198, 362


\bibitem[Vladilo (2001)]{vlad}
Vladilo, G., Centurion, M., Bonifacio, P.  \& Howk, J. C. 2001, 
{\apj}, 557, 1007

\bibitem[Warren (2004)]{swar}
Warren, S. 2004, private communication.

\bibitem[Weingartner \& Draine (2001)]{wda}
Weingartner, J. C. \& Draine, B. T. 2001a, {\apjsupp}, 134, 263

\bibitem[Wolfea\ (2000)]{wolf103} 
Wolfe, A. M., \& Prochaska, J. X. 2000a, {\apj}, 545, 591

\bibitem[Wolfeb\ (2000)]{wolf103} 
Wolfe, A. M., \& Prochaska, J. X. 2000b, {\apj}, 545, 503

\bibitem[Wolfe1\ (2003)]{wolf103} 
Wolfe,  A. M.,  
Prochaska, J. X., \& Gawiser, E. 2003 \apj, 593, 215 (WPG)


\bibitem[Wolfe2\ (200)]{wolf203} 
Wolfe,  A. M., Gawiser, E., 
\& Prochaska, J. X., 2003 \apj, 593, 235 (WGP)

\bibitem[Wolfire et al.\ (1995)]{wolf95} 
Wolfire,  M. G., Hollenbach, D., McKee, C. F., 
Tielens, A. G. G. M.,  \&
Bakes, E. L. O. 1995, \apj, 443,  152 (W95)

\bibitem[Wolfire et al.\ (2002)]{wolf02} 
Wolfire,  M. G., McKee, C. F., Hollenbach, D., 
\& Tielens, A. G. G. M.  
2003, \apj, 587, 278




\end{thebibliography}
\end{document}